\documentclass[preprin,aps,prl,groupedaddress,showpacs,amsmath,amssymb,
floatfix,superscriptaddress]{revtex4}
\usepackage{epsfig}

\begin{document}
\title{\LARGE{Prospects of neutrino oscillation measurements in the
detection of reactor antineutrinos with a medium-baseline experiment
}}
%\author{Mikhail Batygov \\ \and
%  Stephen Dye \\ \and
%  John Learned \\ \and
%  Shigenobu Matsuno \\ \and
%  Sandip Pakvasa \\ \and
%  Gary Varner}

\author{Mikhail Batygov} \email[Email: ]{batygov@phys.hawaii.edu}
\author{Stephen Dye} \email[Email: ] {sdye@hpu.edu}
\author{John Learned} \email[Email: ] {jgl@phys.hawaii.edu}
\author{Shigenobu Matsuno} \email[Email: ] {shige@phys.hawaii.edu}
\author{Sandip Pakvasa} \email[Email: ] {pakvasa@phys.hawaii.edu}
\author{Gary Varner} \email[Email: ] {varner@phys.hawaii.edu}
\affiliation{Department of Physics and Astronomy, University of Hawaii at 
  Manoa, Honolulu, Hawaii 96822, USA}
%\date{October 2008}


\begin{abstract}
Despite the dramatic progress made in neutrino oscillation studies recently, 
several fundamental neutrino parameters remain either unknown or poorly 
measured. 
%Of particular interest are the mixing angle $\theta_{13}$, 
%$\Delta m^2_{13}$, $\Delta m^2_{23}$ and neutrino mass hierarchy, which can 
%only be tackled in new experiments. 
We discuss in detail a method for their measurement
by precision studies of oscillation-caused neutrino energy 
spectrum distortions, for which a large underwater
inverse beta decay detector appears to be a perfect tool. Results determine 
optimal baselines and necessary exposures in the presence of systematic 
uncertainties and the unavoidable background from terrestrial antineutrinos.
%Also discussed are the 
%systematic uncertainties, the unavoidable background from terrestrial
%antienutrinos, optimal baselines, the requirements to the detector and the 
%necessary exposures.
\end{abstract}
\maketitle

\section{Introduction}\label{intro}
Neutrino flavor transformations are determined by the elements of the PMNS 
matrix \cite{Pontecorvo, MNS} and the differences between the squares of 
neutrino mass
eigenvalues. PMNS matrix represents the mixture between flavor and mass 
eigenstates of neutrinos and is conventionally decomposed as shown in
(\ref{PMNS}). Neutrino oscillation experiments can yield the best
estimations for some of those parameters, which has been demonstrated by SNO
\cite {SNO3} and KamLAND \cite {KamLAND3}.

\begin{eqnarray}
U&=&\left[
\begin{array}{ccc}
U_{e 1} & U_{e 2} & U_{e 3} \\
U_{\mu 1} & U_{\mu 2} & U_{\mu 3} \\
U_{\tau 1} & U_{\tau 2} & U_{\tau 3}
\end{array}
\right] \nonumber \\
&=&
\left[
\begin{array}{ccc}
1 & 0 & 0 \\
0 & c_{2 3} & s_{2 3} \\
0 & -s_{2 3} & c_{2 3}
\end{array}
\right]
\left[
\begin{array}{ccc}
c_{1 3} & 0 & s_{1 3}e^{-i\delta} \\
0 & 1 & 0 \\
-s_{1 3}e^{i\delta} & 0 & c_{1 3}
\end{array}
\right] \nonumber \\
&\times&
\left[
\begin{array}{ccc}
c_{1 2} & s_{1 2} & 0 \\
-s_{1 2} & c_{1 2} & 0 \\
0 & 0 & 1
\end{array}
\right]
\left[
\begin{array}{ccc}
e^{i\alpha_1/2} & 0 & 0 \\
0 & e^{i\alpha_2/2} & 0 \\
0 & 0 & 1 \label{PMNS}
\end{array}
\right],
\end{eqnarray}

where $s_{ij} = \sin\theta_{ij}$, $c_{ij} = \cos\theta_{ij}$, $\delta$ is
the phase factor (non-zero if neutrino oscillation violates CP symmetry).
$\alpha_1$ and $\alpha_2$ Majorana phase factors (non-zero only if 
neutrinos are Majorana particles), to which neutrino oscillation experiments 
are not sensitive. 

Besides the PMNS matrix, neutrino oscillations depend on mass eigenvalues or, 
more precisely, on the difference between the squared mass eigenvalues. If 
there are three neutrino mass eigenvalues, then there are only two independent 
differences, the third being either a sum or a difference of the other two. 

Neutrinos studied in experiments are produced in certain flavor eigenstates 
with known abundances of each of them or, as an important special case, in 
only one flavor eigenstate. For example, neutrinos are generated in the
atmosphere with the known
$(\nu_\mu+\bar\nu_\mu)/(\nu_e+\bar\nu_e)$ ratio of about two for low energies; 
solar neutrinos and 
reactor antineutrinos are, initially, all $\nu_e$ and $\bar\nu_e$, 
respectively. 

Detector sensitivity is, generally, flavor dependent. In particular, 
the inverse beta decay, the primary method for detecting 
reactor antineutrinos since the very beginning of neutrino experiments
\cite{ReinesCowan}, involves electron antineutrinos only.
Therefore, the number of detected neutrino events can be different from
the no-oscillation expectation. The deficit of observed neutrinos
compared to no-oscillation prediction was first detected in a solar neutrino
experiment \cite{Davis}. However the rate information alone could not 
provide sufficient evidence to ascribe conclusively the phenomenon of 
neutrino ``disappearance'' to flavor oscillations. 

The energy dependence of neutrino oscillations not only 
changes the neutrino event rate but also distorts the 
observed neutrino energy spectrum. The spectrum distortion provides more 
information about the PNMS matrix components and neutrino mass
eigenstates than rate studies alone can. 

The inverse beta decay method offers excellent energy sensitivity, which is
very valuable for the oscillation studes. Recoil smearing present in this 
reaction is small compared to detector energy resolution, the latter being 
the main limiting factor in the accuracy of $\bar\nu_e$ energy measurement. 
Other advantages include a relatively large cross section of 
the reaction and, most importantly, very powerful background suppression 
due to the characteristic double-coincidence signature. The limitations of 
this method are the $\bar\nu_e$ energy threshold of about 1.8 MeV and weak 
directionality. 

The success of a neutrino oscillation experiment depends not only on the 
characteristics of the detector and on the neutrino source but also on the 
proper choice of the distance between the two (the baseline). 
%A baseline that is too long
%for a parameter in question will make the oscillations on the energy plot,
%especially at the lower part of the spectrum,
%too ``fine'' to be resolved even with the most accurate detectors and, 
%additionally, will decrease the available event statistics. Too short a
%baseline will result in a weak oscillation effect for both the spectrum
%and the overall rate, so that the parameters can not be reliably 
%measured either.
%It is important to note that different oscillation parameters suggest 
%different optimal
%baselines to probe. 
There is no single baseline optimal for all neutrino
oscillation studies. For example, the average baseline of KamLAND experiment,
about 180 km, is fairly good for $\theta_{12}$ and especially for 
$\Delta m^2_{12}$ but not for 
$\theta_{13}$, $\Delta m^2_{13}$ and $\Delta m^2_{23}$. Moreover,
such parameters as detector resolution, the amount, the nature
of the background and the a-priori information about its properties can
affect the optimal baseline value. A tunable baseline
experiment, which implies movable detector or source, may have a big
advantage here.

These considerations, along with the interest in studying terrestrial
antineutrinos, led to the idea of a big KamLAND-like underwater detector
\cite {Hano0, Venice}. The potential of such a detector for neutrino 
oscillation parameter measurements was the primary motivation for the study
presented here. However, the scope of the actual study is much wider and not
limited to the Hanohano project. The results are in fact applicable to any 
similar medium-baselined experiment.

\section{Spectrum distortions due to oscillations}\label{spec_discrimination}

For baselines associated with current and near-future reactor based 
neutrino experiments (up to hundreds of kilometers), the matter effects 
\cite{Wolfenstein, MikheevSmirnov}
critical in solar neutrino studies are not significant, so the 
vacuum oscillation approximation can be used.
As was mentioned above, the inverse beta decay detection is sensitive 
to electron antineutrinos only. Reactors produce exclusively electron 
antineutrinos as well, so the observable effect is the apparent 
``disappearance'' of a fraction of reactor-produced electron antineutrinos. 
The $\bar\nu_e$ ``survival'' probability is given by the formula 
\cite{Schlattl, Bilenky}:

\begin{eqnarray}
P(\bar\nu_e \rightarrow \bar\nu_e)  = 1
&-& \cos^4(\theta_{13})\sin^2(2\theta_{12})\sin^2\Delta_{12} \nonumber \\
&-& \sin^2(2\theta_{13})\cos^2(\theta_{12})\sin^2\Delta_{13} \nonumber \\
&-&\sin^2(2\theta_{13})\sin^2(\theta_{12})\sin^2\Delta_{23}, \label{3flavor}
\end{eqnarray}

where $\Delta_{ij} = \frac{|\Delta m^2_{ij}|R}{4E_\nu}$.
Note that
``atmospheric'' mixing angle $\theta_{23}$ does not affect the $\nu_e$ 
survival and hence not measurable in electron neutrino disappearance 
experiments.  Here, $R$ is the ``baseline'', the distance between the
$\bar\nu_e$ source and the detector.

\begin{figure}[h]
\centering
\scalebox{0.5}{\includegraphics{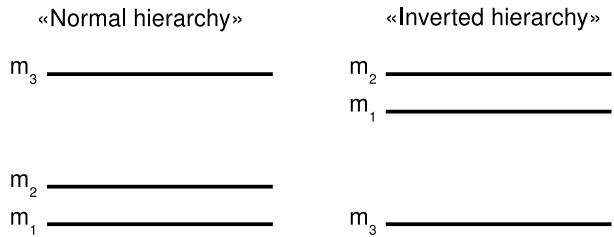}}
\caption{Two neutrino mass hierarchies allowable with currently
available data; all three combinations with $m_2 < m_1$ are
excluded by solar experiments; the $m_1 < m_3 < m_2$ order is excluded
by the fact that $\Delta m^2_{12} \ll \Delta m^2_{23} \approx 
\Delta m^2_{13}$.}
\label{hierarchies}
\end{figure}

Given the evidence from solar neutrino experiments \cite{SNO, SK-solar} 
that $m_2 > m_1$, and the knowledge
that $\Delta m^2_{23} \gg \Delta m^2_{12}$ from SuperK \cite{SuperK2004}, K2K
\cite{K2K}, MINOS \cite{MINOS} on the one hand and KamLAND \cite{KamLAND3} on 
the other, only two neutrino hierarchies out of possible six are allowed with 
currently available data. They are
commonly referred to as ``Normal Hierarchy, NH'' ($m_1 < m_2 < m_3$) and 
``Inverted Hierarchy, IH'' ($m_3 < m_1 < m_2$) (Fig. \ref{hierarchies}). 
The former implies that
$\Delta m^2_{13} > \Delta m^2_{23}$, the latter that 
$\Delta m^2_{23} > \Delta m^2_{13}$, so the sufficiently precise measurement 
of the those squared mass differences should be enough to establish the
neutrino mass hierarchy. 

The measurement of $\Delta m^2_{13}$, $\Delta m^2_{23}$ and the mass 
hierarchy with this approach is possible, in theory, only if 
$\theta_{13}$ is finite and, in practice, if this
mixing angle is large enough. Moreover, if the ``solar'' mixing is maximum 
($\theta_{12}=\pi/4$), the $\Delta m^2_{13}$ and $\Delta m^2_{23}$ 
become mutually indistinguishable, thus still ruling out the mass 
hierarchy study, although their values may still be determined without knowing 
``which is which''. This maximum mixing is strongly disfavored by 
KamLAND \cite{KamLAND3} and essentially excluded by solar eperiments 
\cite{SNO, SK-solar}. 
Unfortunately, the same can not be said about the $\theta_{13}$ since only
the upper limit for this value exists today and there is no experimental 
evidence that it is not zero. If it is, then future $\bar\nu_e$ vacuum 
oscillation experiments are limited to probing $\theta_{12}$ and 
$\Delta m^2_{12}$ along with setting still better upper limits on the
$\theta_{13}$ itself. That said, global analysis shows a slight preference
for non-zero $\theta_{13}$ \cite{ItalianSpanish}.

\begin{figure}[h]
\centering
\scalebox{0.7}{\includegraphics{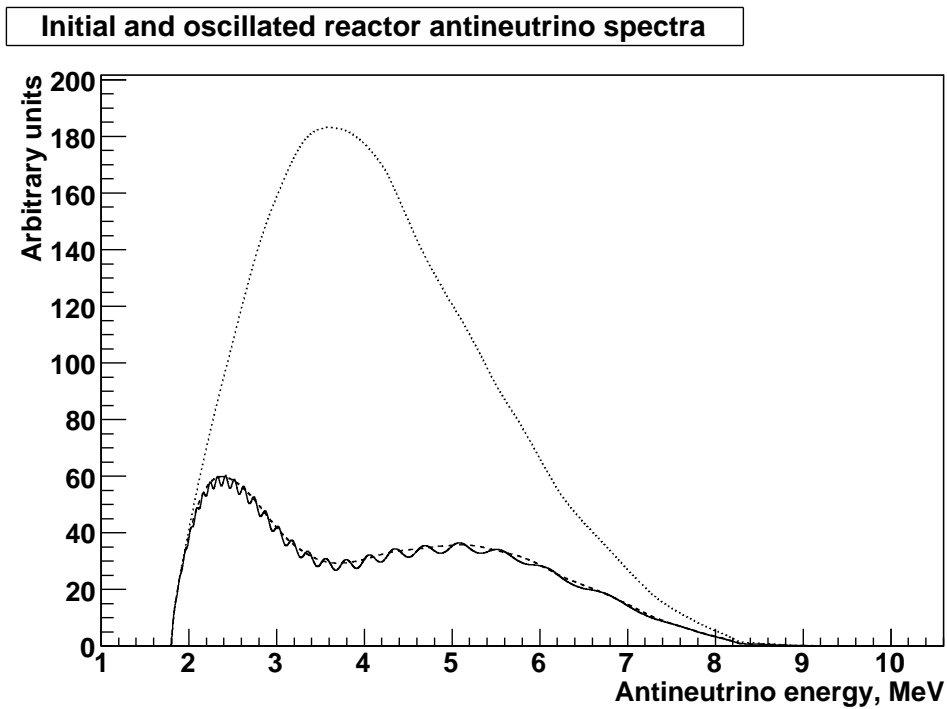}}
\caption{Typical reactor $\bar\nu_e$ spectrum: non-oscillated (dotted), 
with $\theta_{13} = 0$ (dashed), and with $\theta_{13} = 0.05$ (solid).
}
\label{raw}
\end{figure}

A typical reactor $\bar\nu_e$ spectrum \cite{VogelReactor, Schreckenbach} 
multiplied by the inverse beta decay
cross section \cite{Vogel} is shown in Figure \ref{raw}, dotted. 
The antineutrinos are generated in $\beta^-$ decays of short
living fission products of initial fissionable fuel isotopes: $^{235}U$,
$^{238}U$, $^{239}Pu$, $^{241}Pu$. For this study, the ratio of the isotopes 
is taken the same as in \cite{KamLAND3}. Such a spectrum can be 
observed at very short-baselined experiments (baseline $\ll$ 1 km),
where oscillation effects are negligible. 

At much longer baselines (30 km and above), the ``solar'' oscillations 
governed by $\theta_{12}$ and $\Delta m^2_{12}$
lead to an energy-dependent deficit of the observed $\bar\nu_e$ events. The
effect of those oscillations alone 
%(which is the real case if $\theta_{13}$ happens to be zero) 
is the ``coarse'' oscillatory pattern of event 
deficit over the spectrum (Figure \ref{raw}, dashed) with a high amplitude 
(determined by $\sin^2\theta_{12}$) and a relatively low frequency 
(determined by $\Delta m^2_{12}$).

The amplitude of oscillations driven by the squared mass differences 
$\Delta m^2_{13}$ and $\Delta m^2_{23}$ is proportional to 
$\sin^22\theta_{13}$ and much smaller than that of ``solar'' oscillations 
for any currently allowed value of this mixing angle.
Because $\Delta m^2_{13}$ and $\Delta m^2_{23}$ are known to be 
larger than $\Delta m^2_{12}$, the frequency of those
sub-dominant oscillations is higher. A typical $\bar\nu_e$ energy spectrum
expected for a non-zero $\theta_{13}$ is shown in Figure \ref{raw}, solid.

The spectrum analysis approach has already been successfully
used by KamLAND to set by now the best limits on $\Delta m^2_{12}$
and to confirm SNO and SuperK values for $\theta_{12}$. The idea to measure
the remaining three of the five oscillation parameters by precision
measurement of the sub-dominant oscillation pattern in a reactor $\bar\nu_e$ 
disappearance experiment has been already suggested and thoroughly examined
\cite {Petcov1, Petcov2, Petcov3}. 

In this paper, we examine the capacity of an intermeidate baseline (30-90 km)
reactor $\bar\nu_e$ experiment for measuring $\theta_{12}$, 
$\theta_{13}$, $\Delta m^2_{12}$, $\Delta m^2_{13}$, $\Delta m^2_{23}$ 
and neutrino mass hierarchy. Although this study has been motivated by the 
project of a big underwater detector Hanohano (\cite {Hano0, Venice}), we make
no assertions specific for that particular choice. A special emphasis is placed
on the systematic uncertainties and technical
limitations present in any real experiment. In our study of the sensitivity 
to each of the oscillation parameters we take into account the impact of 
those uncertainties, as well as some detector parameters and the baselines on 
the resulting performance to formulate in a quantitative way the requirements 
to which such an experiment must conform.

\section{The scope of analysis}\label{analysis-scope}

In this study, we consider the measurement of all the oscillation parameters 
to which such $\bar\nu_e$ disappearance experiments are sensitive:
% None of 
% the oscillation or geo-neutrino parameters is known and expected to be known 
% any time soon with an accuracy comparable to what the future experi itself 
% is targeting 
% (the mixing angle $\theta_{13}$ being the only possible exception). 
% Because of that, 
%$\theta_{12}$, $\theta_{13}$, $\Delta m^2_{12}$, 
%$\Delta m^2_{13}$ are left free to vary within currently allowable 
%regions. 
$\theta_{12}$, $\theta_{13}$, $\Delta m^2_{12}$, 
$\Delta m^2_{13}$, $\Delta m^2_{23}$. Three types of detector-related 
systematic uncertainties 
are considered which are present to some extent in any experiment and 
are capable of a non-trivial impact on the sensitivity to the target 
parameters. Although the success of the Borexino experiment \cite{Borexino} 
suggests that careful detector design can make the inverse-beta 
based $\bar\nu_e$ detection 
almost background-free, geologically
produced antineutrinos \cite{KLGeo, KamLAND3} will technically
remain a background source for a reactor $\bar\nu_e$ study in the lower
energy region. What makes this background especially significant is the lack
of exact information about its overall intensity and the relative amounts
of antineutrinos produced in the ``Uranium-Radium'' and ``Thorium'' decay
series. This amounts to two more systematic uncertainties which 
have to be left unconstrained within geologically feasible models.

%Ignoring this background
%and the associated uncertainties would produce some over-optimistic 
%estimations of the sensitivity, especially that to the ``solar'' oscillation 
%parameters, and may even provide wrong directions to the optimum
%setup of the experiment.

The following detector-related uncertainties were accounted for:
\begin{itemize}
\item
  The uncertainty in the predicted event rate. It is sensitive to the number of 
target protons (due to fiducial volume estimation error and uncertainty of the
scintillator composition), the efficiency of coincidence
selection cuts, and live time estimation error. For current similar 
experiments, this error tends to be on the order of 1 to 5\%. Below we 
refer to this as ``efficiency'' error.
\item
  The uncertainty in the detector energy resolution estimation. Although often 
ignored in current experiments, it can have a considerable effect on the 
measurement of the $\theta_{13}$ mixing angle from medium baselines. 
Numerically, it can be quite big (about 10\%) depending on the detector 
calibration options.
\item
  The ``linear'' energy scale uncertainty. This is the 
uncertainty in the average number of photoelectrons produced by an 1 MeV 
event. The amount of this uncertainty depends on the detector calibration 
as well. Normally it
can be made quite small (around 1\%) but its impact on the resulting 
accuracy of the parameter estimation may still be noticeable.
\end{itemize}

The energy scale in scintillator-based detectors is in fact substantially
non-linear, this non-linearity always producing additional systematic
uncertainties which are often rather tricky to parametrize. However
the study of this error is very detector-specific, requires extensive
Monte-Carlo simulations with real calibration data feedback and considering 
it at this stage would be too speculative. Although, KamLAND internal studies
indicate that the nonlinear energy scale uncertainty is less of an issue than
the linear one which we can take into account now, Hanohano or any other
future experiment will have to revisit this issue, once the real experimental 
feedback from the detector becomes available.

The geo-neutrinos yield two more systematic uncertainties:
\begin{itemize}
\item
  Total detectable terrestrial antineutrino flux, conventionally expressed
in Terrestrial Neutrino Units (TNU) defined as the number of inverse-beta
decay interactions per $10^{32}$ free protons per year. 
\item
  The ratio of $\bar\nu_e$ originating from the $^{238}U$ decay chain to
those coming from the $^{232}Th$ decay chain. 
\end{itemize}

Although geological models do provide some guidelines for the expected 
geo-neutrino
flux and KamLAND was able to produce the first experimental measurement
of the flux, these data are of little use for the 
purpose of future experiments, including Hanohano, because the geo-neutrino 
estimation precision needed to produce
an appreciable advantage over the ``agnostic'' approach is about one order 
of magnitude higher than available now. 

%Actually Hanohano itself can yield a
%much better estimation for the geo-neutrino flux already during the neutrino
%oscillation phase of the experiment, even despite the reactor $\bar\nu_e$ 
%background to the terrestrial antineutrinos. Because of that, we made a quick 
%assessment of the accuracy with which the terrestrial $\bar\nu_e$ flux 
%parameter estimation can be improved in the presence of the reactor background.

\section{Statistical analysis procedure}\label{analysis}

Since we've included background and systematics in the analysis, the direct 
likelihood approach has been chosen 
over the combination of the matched digital filter and the Fourier transform 
of the spectrum employed in the earlier publications dedicated to or motivated
by Hanohano project \cite{Hano1, Venice}. This approach facilitates the 
accommodation of
the systematic uncertainties and the background. The likelihood method used 
here is the unbinned statistical analysis similar to the one employed by 
KamLAND experiment 
\cite{KamLAND1, KamLAND2, KLGeo, KamLAND3}. Instead of the real experimental 
data, a series of
``experiments'' can be simulated as sequences of ``events'' with energies 
distributed according to the spectra distorted by different oscillation 
parameters (including the background). The potential sensitivity is 
essentially the ability of the data analysis to distinguish between 
different hypotheses about the oscillation parameter sets.

This study is based on the ``rate+shape'' likelihood function defined for a 
real or simulated experiment as:

\begin{equation}\label{likelihood}
L(\vec{E}_{\bar\nu_e}|\vec{\eta}) = 
e^{-N_{exp}}\prod_{i=1}^{N_{events}}f(E^i_{\bar\nu_e}|\vec{\eta}),
\end{equation}

where $\vec{E}_{\bar\nu_e} = \{E^1_{\bar\nu_e}...E^N_{\bar\nu_e}\}$ are the 
event energies, N --- the
number of observed events, $N_{exp}(\vec{\eta})$ --- the expected number of 
events (given the set of parameters $\vec\eta \equiv \{\Delta m^2_{12},
\Delta m^2_{13}, \Delta m^2_{23}, \theta_{12}, \theta_{13}\}$), 
$f(E^i_{\bar\nu_e}|\vec{\eta})$ --- $\bar\nu_e$ energy spectrum normalized 
to $N_{exp}$ (after the distortion by the set of parameters 
$\vec\eta$). Note that while both $\vec{E}_{\bar\nu_e}$ and 
$\vec{\eta}$ are denoted as vectors, these vectors are in different 
spaces. The $\vec{E}_{\bar\nu_e}$ has as many dimensions as the sum of
the number of $\bar\nu_e$ and background events in the experiment, and is 
fixed for a given experiment. The $\vec{\eta}$ lies in the parameter space, 
its dimensionality being the number of unknown parameters to be fitted, and 
is variable. 

The best fit is obtained by varying the parameter vector $\vec\eta$ to
achieve the maximum value of $L$ or its logarithm, the latter often being 
more convenient to calculate and handle. After the best fit point 
$\vec\eta^0$ has been found, the general prescription to
evaluate the sensitivity to some individual parameter $\eta_k$ is the 
following:
\begin{itemize}
\item 
  Make a small increment (or decrement) $\epsilon$ to the $\eta_k$ from 
the ``best fit'' point $\eta_k^0$: $\eta'_k = \eta_k^0+\epsilon$.
\item
  Find a new point of maximum likelihood by varying all the parameters
$\vec\eta$ except for $\eta_k$ which is kept fixed at $\eta'_k$. The new
maximum $L'$ over the subspace constrained by the requirement 
$\eta_k = \eta'_k$ is not higher than the global maximum $L^0$.
\item
  Repeat the above steps with varying $\epsilon$ until the condition 
$\log L^0 - log L' = Q_{CL}$ is met for both positive and negative 
increments of $\epsilon$. The corresponding points $\eta'^{low}_k$ and 
$\eta'^{high}_k$ will limit the confidence range for the k-th parameter. 
\end{itemize}

The value $Q_{CL}$ depends on the confidence level for which the range
is to be determined. For an individual parameter variation and the 
confidence level equal to $1 \sigma$, $Q_1 = \frac{1}{2}$. In general,
for a CL of $n\sigma$, $Q_n=\frac{1}{2}n^2$. When instead of a 
one-dimensional confidence range, a multidimensional confidence
region in the parameter subspace is required, the 
values $Q_{CL}$ will be different but the general procedure will not change. 
The same is true for the case of discriminating two discrete hypotheses,
e.g. between the normal and the inverted neutrino mass hierarchies. More 
detailed information on the likelihood analysis can be found in 
\cite{GlenCowan}.

For a simulated experiment, the experimental
points $\vec E_{\bar\nu_e}$ do not exist in the first place and are 
generated according to some reasonable choice of parameters $\vec\eta$. 
Except for this initial stage, the rest of the analysis is the 
same as described above. If the initial choice of the parameters
to simulate the events is not too far off, this procedure will yield
an accurate prediction for the sensitivity of the actual experiment.

Systematic uncertainties are introduced by adding ``hidden'' parameters to
the parameter space and allowing them to vary during the search for the 
maximum likelihood as well. When some information about these values is 
available
a ``penalty'' term is subtracted from $\log(L)$ to account for the fact 
that big deviations from the central values of those hidden parameters are
unlikely. If the parameters of uncertainty are normally
distributed around their central value and if all systematic uncertainty 
parameters are uncorrelated, the ``penalty'' term takes on the form:

\begin{equation}\label{penalty-term}
\frac{1}{2}\sum_{j=1}^{N_{SP}}\frac{\delta\eta^2_{k_{j}}}{\sigma^2_j},
\end{equation}

where $N_{SP}$ is the number of systematic uncertainty parameters, $k_j$
is the index of the parameter corresponding to the $j$-th uncertainty,
$\delta\eta_{k_{j}}$ is the deviation of the $k_j$-th parameter from
its most probable value, and $\sigma_j$ is the value of systematic error 
ascribed to the $j$-th uncertainty. When some uncertainties are correlated,
the penalty term becomes a more general positive definite quadratic form
but for our current study that is not the case.

The full equation for the likelihood logarithm with systematic 
uncertainties takes on the form:

\begin{equation}\label{full-log-likelihood}
\log L(\vec{E}_{\bar\nu_e}|\vec{\eta}) = 
-N_{exp}+\sum_{i=1}^{N_{events}}\log(f(E^i_{\bar\nu_e}|\vec{\eta}))
+
\frac{1}{2}\sum_{j=1}^{N_{SP}}\frac{\delta\eta^2_{k_{j}}}{\sigma^2_j}
\end{equation}

In this work, we used the total of nine continuous parameters.
These are four neutrino oscillation parameters: $\sin^2(2\theta_{12})$, 
$\sin^2(2\theta_{13})$, $\Delta m^2_{12}$, $\Delta m^2_{13}$. Note that 
for each of the two mass hierarchies, $\Delta m^2_{23}$ is determined by 
the two other squared mass differences and has not to be introduced into 
the parameter space. Five parameters were dedicated to systematic
uncertainties: two geo-neutrino parameters and three detector-associated
systematic errors as described in the previous section. The geo-neutrino 
parameters are
left unconstrained, which is equivalent to infinite $\sigma$ in 
(\ref{penalty-term}). The default values for the systematic errors in
``efficiency'', energy resolution estimation and energy scale were taken
to be 2\%, 8\% and 1\%, respectively, which is reasonably conservative
for experiments of this kind. Additionally, two extreme cases 
were analyzed: the most ``optimistic'' one --- with the corresponding 
parameters 
fixed at zero deviations as if they were known exactly, and the most
``pessimistic'' one --- with those three uncertainties left unconstrained 
as well as geo-neutrino parameters. Although practically impossible, these
limiting cases indicate how much sensitivity can be gained by
improving the systematics and, conversely, how much would be lost
if the systematic errors of the real experiment happen to be worse
than expected.

\section{``Solar'' mixing angle $\theta_{12}$}

The ``solar'' mixing angle has been fairly well constrained by SNO 
\cite{SNO3} and 
KamLAND \cite{KamLAND3}. Our computations suggest that there is still
an opportunity for a significant improvement, though. 

\begin{figure}[p]
\centering
\scalebox{0.7}{\includegraphics{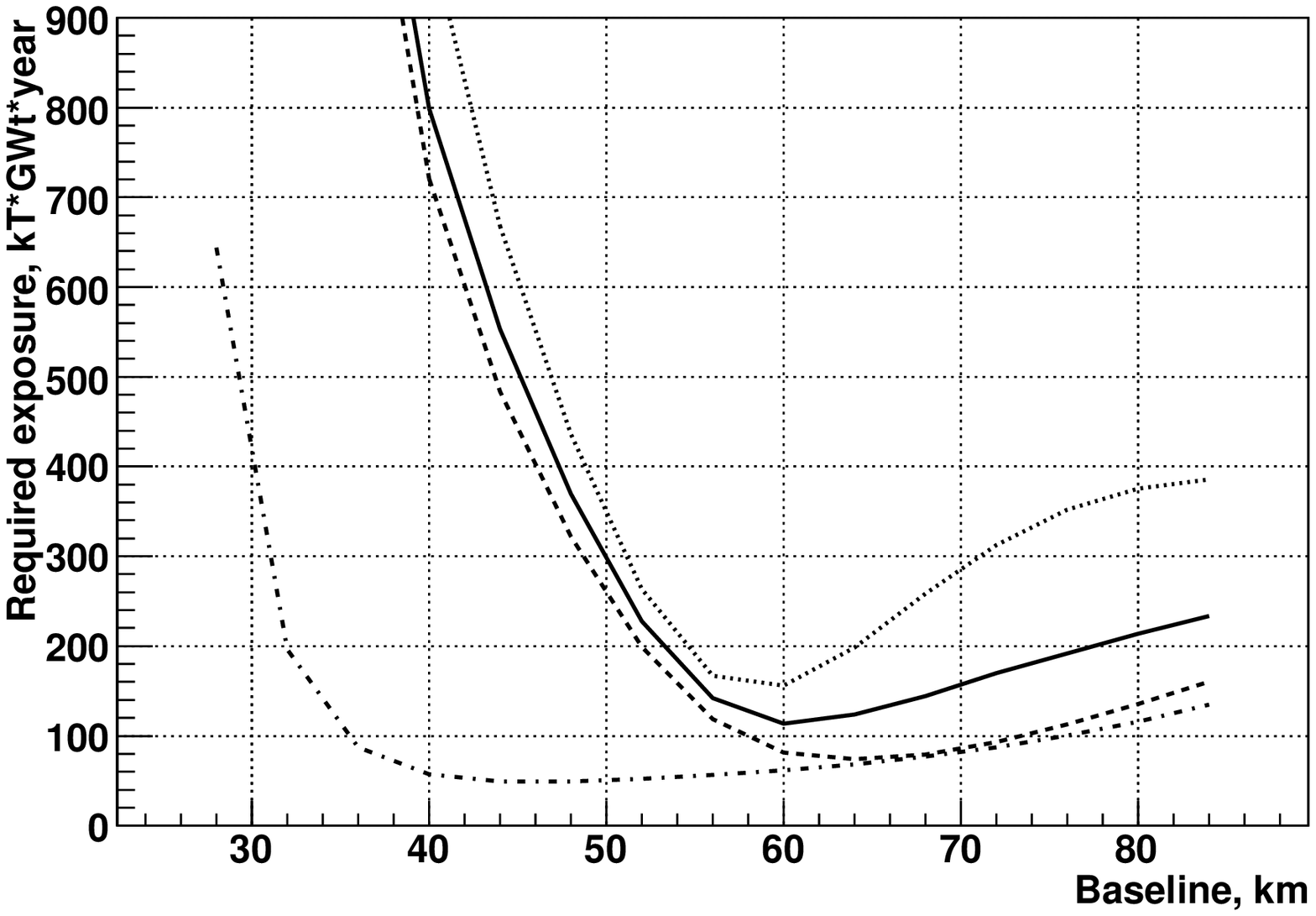}}
\caption{Exposure yielding the sensitivity of 0.01 in $\sin^22\theta_{12}$,
as a function of baseline: with unconstrained detector systematics (dotted),
with ``default'' detector systematics (solid), assuming no detector 
systematics (dashed), assuming no detector systematics and no geo-neutrinos
(dot-dashed).}
\label{ss12}
\end{figure}

This measurement is moderately sensitive to detector-based systematic 
uncertainties but the
terrestrial antineutrino background is much more troublesome (Fig \ref{ss12}). 
These background decrease the sensitivity by about a factor of two and drives
the optimum baseline for this parameter to about 60 km, which conflicts
with the goal of measuring $\theta_{13}$ and $\Delta m^2_{13}$
in the same experiment as well. Constraining the geo-neutrino flux
would change the situation, but currently
there seems to be no way of doing that. The geo-neutrino flux measurements
made by KamLAND \cite{KLGeo, KamLAND3} are not nearly 
precise enough to improve the situation noticeably and, besides, not
directly applicable to a future experiment located elsewhere, especially in 
the ocean.

Even with that background and the associated uncertainty, the sensitivity 
of the medium-baseline experiment is noteworthy. For a 10 KT detector 
at 50 km from a 6 GWt nuclear plant, an exposure of
300 gigawatt-kiloton-years (5 years) is required to achieve the one-sigma 
confidence range of 0.01 in $sin^2(2\theta_{12})$, which is about 5 times 
better than the current best estimation. At 60 km, the same precision can
be achieved with just above one-third of this exposure.

Although higher energy resolution is always better, the $\theta_{12}$ study 
does not exhibit appreciable dependence on this parameter and 
$0.05\times\sqrt{E_{vis}[MeV]}$ is almost as good as 
$0.025\times\sqrt{E_{vis}[MeV]}$. Of the detector-associated systematics, 
the most significant is the ``efficiency'' uncertainty.

\section{``Solar'' squared mass difference $\Delta m^2_{12}$}

The measurement of this parameter by KamLAND is more
difficult to improve on. For example, the target sensitivity for Hanohano 
is $0.07\times10^{-5}eV^2$, which would be about three times better than 
the current best estimation. As our calculations suggest, this can be achieved 
in 300 gigawatt-kiloton-years at the 60 km baseline or in 450
gigawatt-kiloton years from 50 km. 

\begin{figure}[p]
\centering
\scalebox{0.7}{\includegraphics{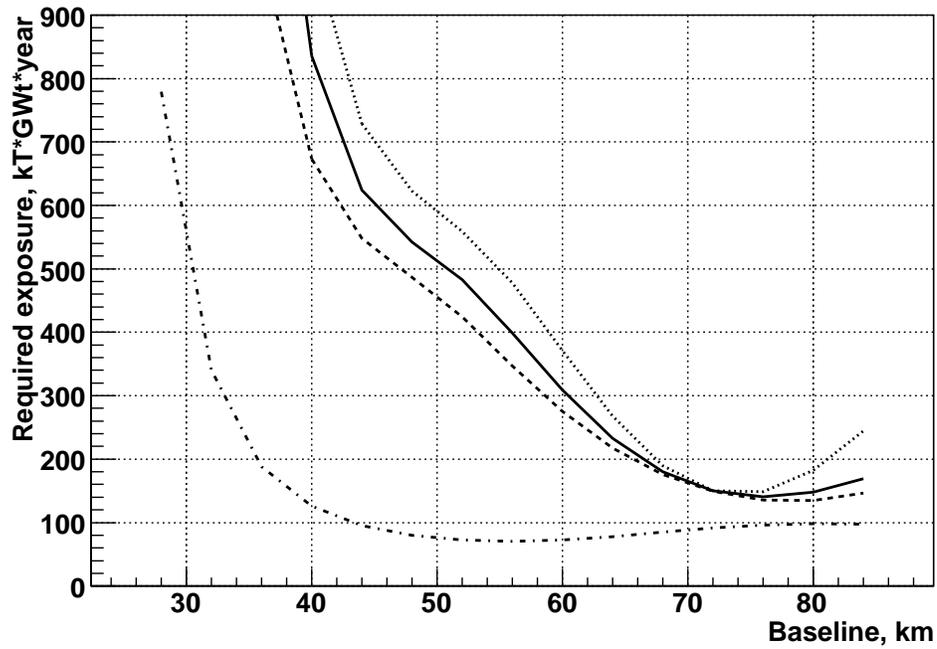}}
\caption{Exposure yielding the sensitivity of $0.07\times 10^{-5} eV^2$ in  
$\Delta m^2_{12}$, as a function of baseline: with unconstrained 
detector systematics (dotted), with ``default'' detector systematics 
(solid), assuming no detector systematics (dashed), assuming no detector 
systematics and no geo-neutrinos (dot-dashed).}
\label{dms12}
\end{figure}

Still longer baselines offer better sensitivity for this particular study 
(Fig \ref{dms12}) but would be clearly sub-optimal for all
other oscillation parameters. The part (b) of the plot shows
the dramatic effect of terrestrial neutrino background and, particularly, the 
uncertainty of this background. Without geo-neutinos, the
same detector would be four times more efficient at 60 km and seven 
times at 50 km.

The fact that geo-neutrinos drive the optimum baseline towards longer 
distance may seem somewhat counter-intuitive. The shorter the baseline, the
higher the reactor $\bar\nu_e$ rate, so the relative fraction of 
terrestrial $\bar\nu_e$ background is smaller and should have a smaller 
effect. However, at shorter baselines, the reactor $\bar\nu_e$ deficit
due to oscillations appears mostly in the lower-energy zone where it is 
harder to separate from the variation in the terrestrial neutrino background. 
%(Fig \ref{dms12-geo})

Like the $\theta_{12}$ measurement, this study is not demanding of
detector energy resolution and not particularly sensitive to 
detector-associated systematics.

\section{Mixing angle $\theta_{13}$}

This is a very important oscillation parameter not only because of its 
theoretical significance but also because its value defines the amplitude
of the sub-dominant high-frequency oscillations governed by $\Delta m_{13}$ 
and $\Delta m_{23}$. Only if $\theta_{13}$ is not zero (and not too small) is
it possible to measure those mass squared differences in $\bar\nu_e$
disappearance experiments. Currently, only an upper bound for
this angle is known (from the CHOOZ experiment \cite{CHOOZ}): 
$\sin^22\theta < 0.1$. 
Several experiments are proposed or already under construction to set
better limits and the Hanohano detector can 
contribute to those efforts. 

\begin{figure}[p]
\centering
\scalebox{0.7}{\includegraphics{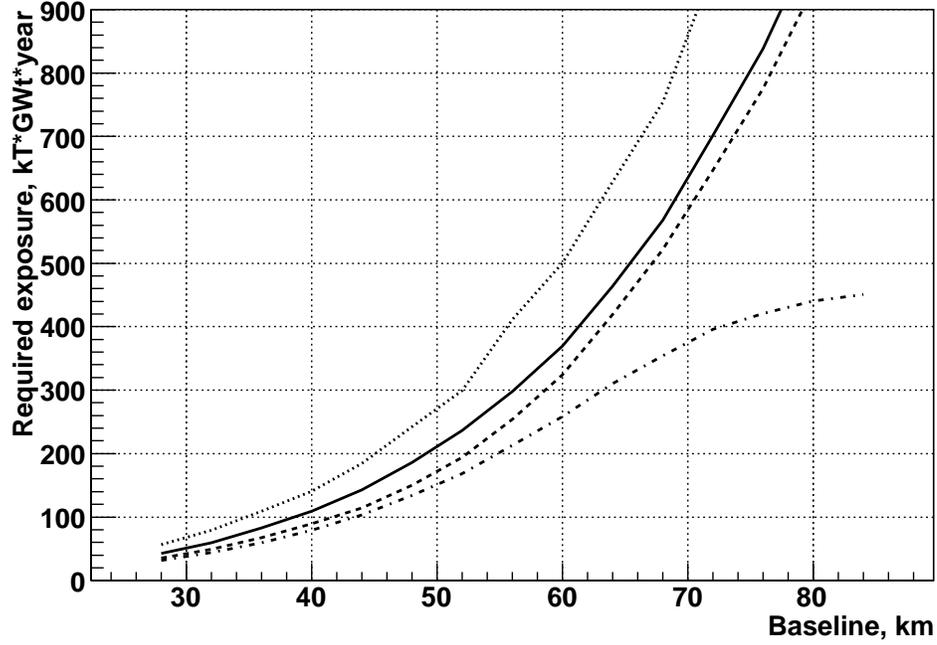}}
\caption{Exposure yielding the sensitivity of 0.02 in $\sin^22\theta_{13}$,
as a function of baseline: with unconstrained detector systematics 
(dotted), with ``default'' detector systematics (solid), assuming no detector 
systematics (dashed), assuming no detector systematics and no geo-neutrinos
(dot-dashed).
Detector energy resolution equal to $0.025\times\sqrt{E_{vis}[MeV]}$ is 
assumed.}
\label{ss13}
\end{figure}

The sensitivity profiles (Figure \ref{ss13}) show that medium
baselines (above 30 km) are not optimal for this study and much shorter 
ones are better from the statistical standpoint. However even at 50 km
the absolute sensitivity can be quite impressive with a big
detector. 
%Hanohano can get to 0.02 in $\sin^22\theta$ in about
%200 gigawatt-kiloton-years, provided it
%has a really good energy resolution. 
Except for the longest baselines (60 km and above) which are clearly 
sub-optimal, this study is not
severly affected by the geo-neutrino background and its uncertainties. 
The systematics of the detector itself, however, play a more important
role here. At 50 km, the main systematic error is the uncertainty
of energy resolution estimation, followed by the ``efficiency'' error. 
At shorter baselines the ``efficiency'' uncertainty dominates. 

Although the medium baselines have a strong statistical disadvantage
for $\theta_{13}$ measurement, they also have the compelling feature that 
systematic uncertainties do not ruin the measurement. Unlike the
shorter baseline experiments where relatively more information is
obtained through the neutrino event rate, the spectrum shape distortion
characteristic of medium baseline is not so easy to imitate by
any of the detector systematic errors. This means that, in the
long run when even better accuracy for $\theta_{13}$ is required, medium 
baseline experiments may prove to be more robust.

\begin{figure}[p]
\centering
\scalebox{0.7}{\includegraphics{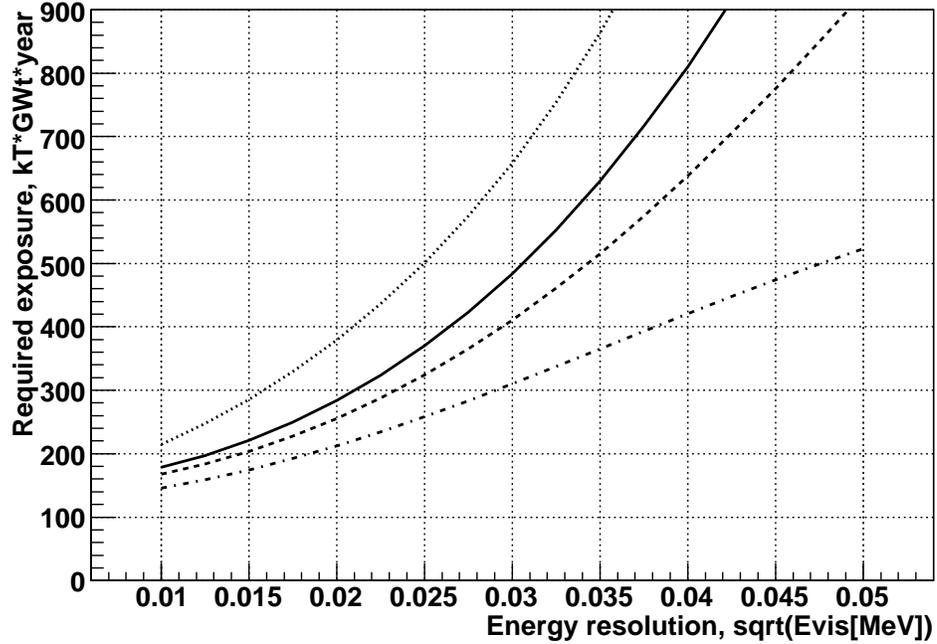}}
\caption{Exposure yielding the sensitivity of 0.02 in
$\sin^22\theta_{13}$ from the baseline of 60 km, as a function of the 
detector energy resolution: with unconstrained detector systematics 
(dotted), with ``default'' detector systematics (solid), assuming no detector 
systematics (dashed), assuming no detector systematics and no geo-neutrinos
(dot-dashed).}
\label{ss13R}
\end{figure}

Figure \ref{ss13R} exhibits another important feature of this measurement:
its energy resolution dependence. Although 
not as critical as for the hierarchy study (below), the effect of
detector energy resolution is quite noticeable. Compromising this parameter
to $0.05\times\sqrt{E_{vis}[MeV]}$ from the 
$0.025\times\sqrt{E_{vis}[MeV]}$ (as projected for Hanohano) will cost about
2.5 times the exposure.

\section{$\Delta m^2_{13}$ and $\Delta m^2_{23}$}

Unlike all previously considered parameters where the potential sensitivity 
of an experiment could be predicted more or less accurately based just 
on the $\bar\nu_e$ exposure, this measurement depends on the
value of $\theta_{13}$ which is still unknown. Any quantitative
sensitivity prediction makes sense only with some particular value of 
$\theta_{13}$ in mind. The larger the mixing angle, the easier it is to 
determine 
$\Delta m^2_{13}$, $\Delta m^2_{23}$ and neutrino mass hierarchy. It has
been found that the sensitivity scales approximately as the square of 
$\sin^22\theta_{13}$. 
In other words, getting the same sensitivity in 
$\Delta m^2_{13}$, $\Delta m^2_{23}$ and neutrino mass hierarchy if 
$\sin^22\theta_{13} = 0.01$ 
will take four times the exposure required if $\sin^22\theta_{13} = 0.02$.
In this paper we carried out calculations for two scenarios: 
$\sin^22\theta_{13} = 0.05$, and $\sin^22\theta_{13} = 0.025$. 

Another ambiguity associated with the study of $\Delta m^2_{13}$ and
$\Delta m^2_{23}$ follows from the closely related question of neutrino 
mass hierarchy. Depending on the actual value of the mixing angle 
$\theta_{13}$, the neutrino mass hierarchy may turn out unfeasible to 
establish at an adequate CL. At the same time, within each of the two 
possible hierarchies, stringent limits on both $\Delta m^2_{13}$ and 
$\Delta m^2_{23}$ may still be set with reasonable exposures. 
Since the ambiguity is at worst only a two-fold one,
it makes sense to estimate the ``known-hierarchy'' sensitivity to 
either $\Delta m^2_{13}$ or $\Delta m^2_{23}$.

\begin{figure}[p]
\centering
\scalebox{0.7}{\includegraphics{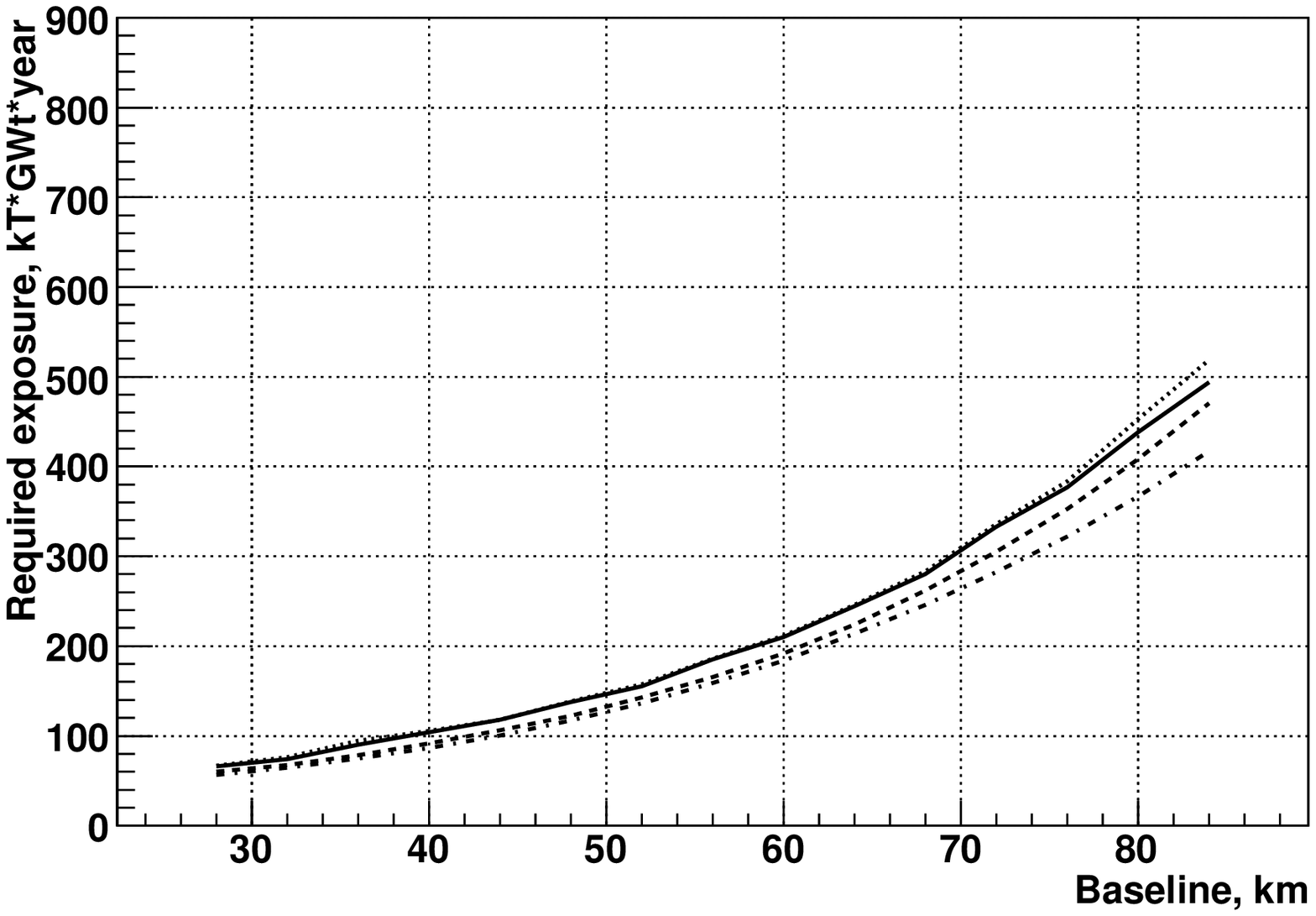}}
\caption{\textit {Assuming $\sin^22\theta_{13} = 0.05$:} 
exposure yielding the sensitivity of $0.025\times 10^{-3} eV^2$ in 
$\Delta m^2_{13}$ within a given mass hierarchy,
as a function of baseline: with unconstrained detector systematics 
(dotted), with ``default'' detector systematics (solid), assuming no detector 
systematics (dashed), assuming no detector systematics and no geo-neutrinos
(dot-dashed). Detector energy resolution equal to 
$0.025\times\sqrt{E_{vis}[MeV]}$ is assumed.}
\label{dms13-50}
\end{figure}

\begin{figure}[p]
\centering
\scalebox{0.7}{\includegraphics{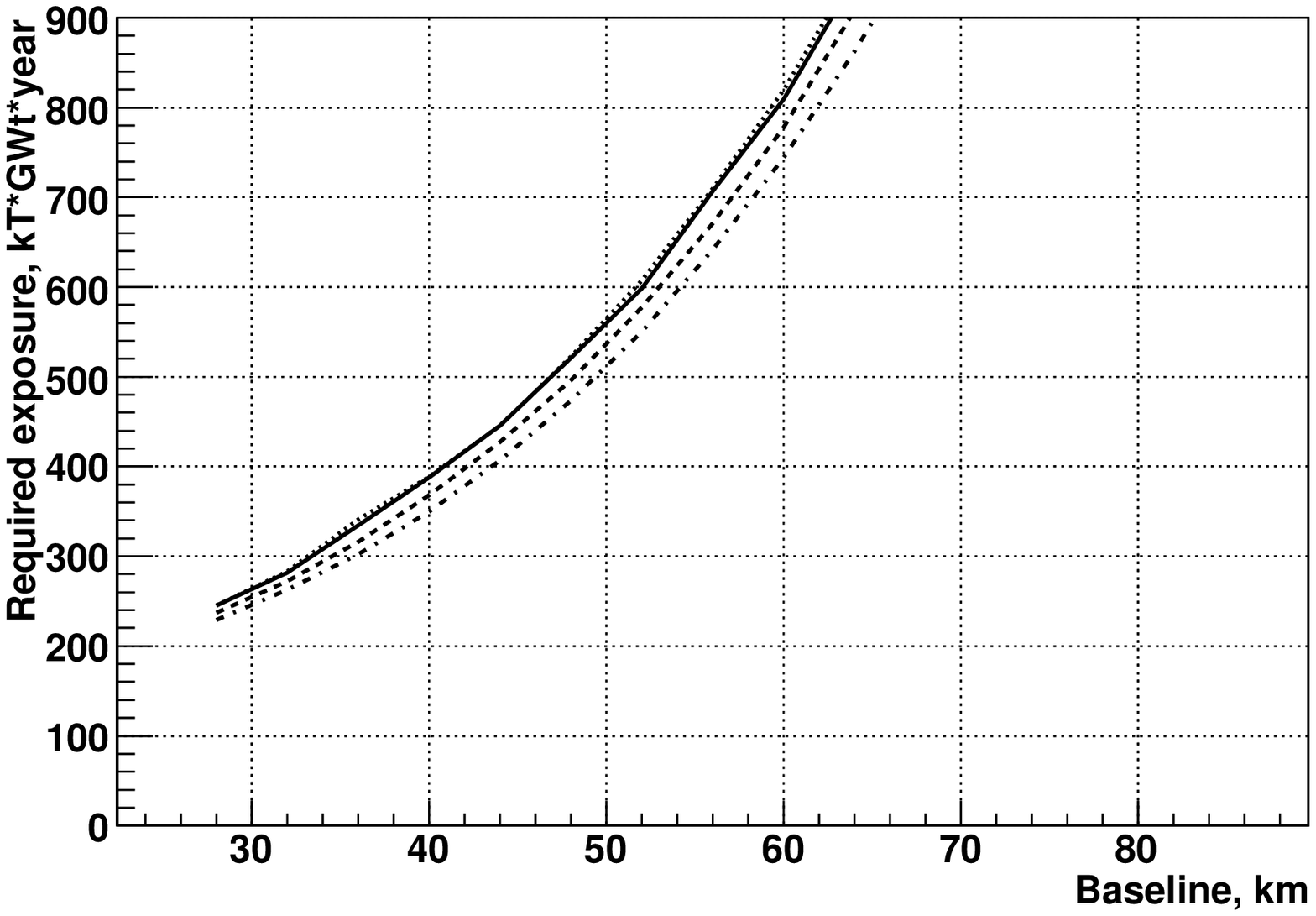}}
\caption{\textit {Assuming $\sin^22\theta_{13} = 0.025$:}
exposure yielding the sensitivity of $0.025\times 10^{-3} eV^2$ in 
$\Delta m^2_{13}$ within a given mass hierarchy, as a function of the 
baseline: with unconstrained detector systematics 
(dotted), with ``default'' detector systematics (solid), assuming no detector 
systematics (dashed), assuming no detector systematics and no geo-neutrinos
(dot-dashed). Detector energy resolution equal to 
$0.025\times\sqrt{E_{vis}[MeV]}$ is assumed.}
\label{dms13-25}
\end{figure}

\begin{figure}[p]
\centering
\scalebox{0.7}{\includegraphics{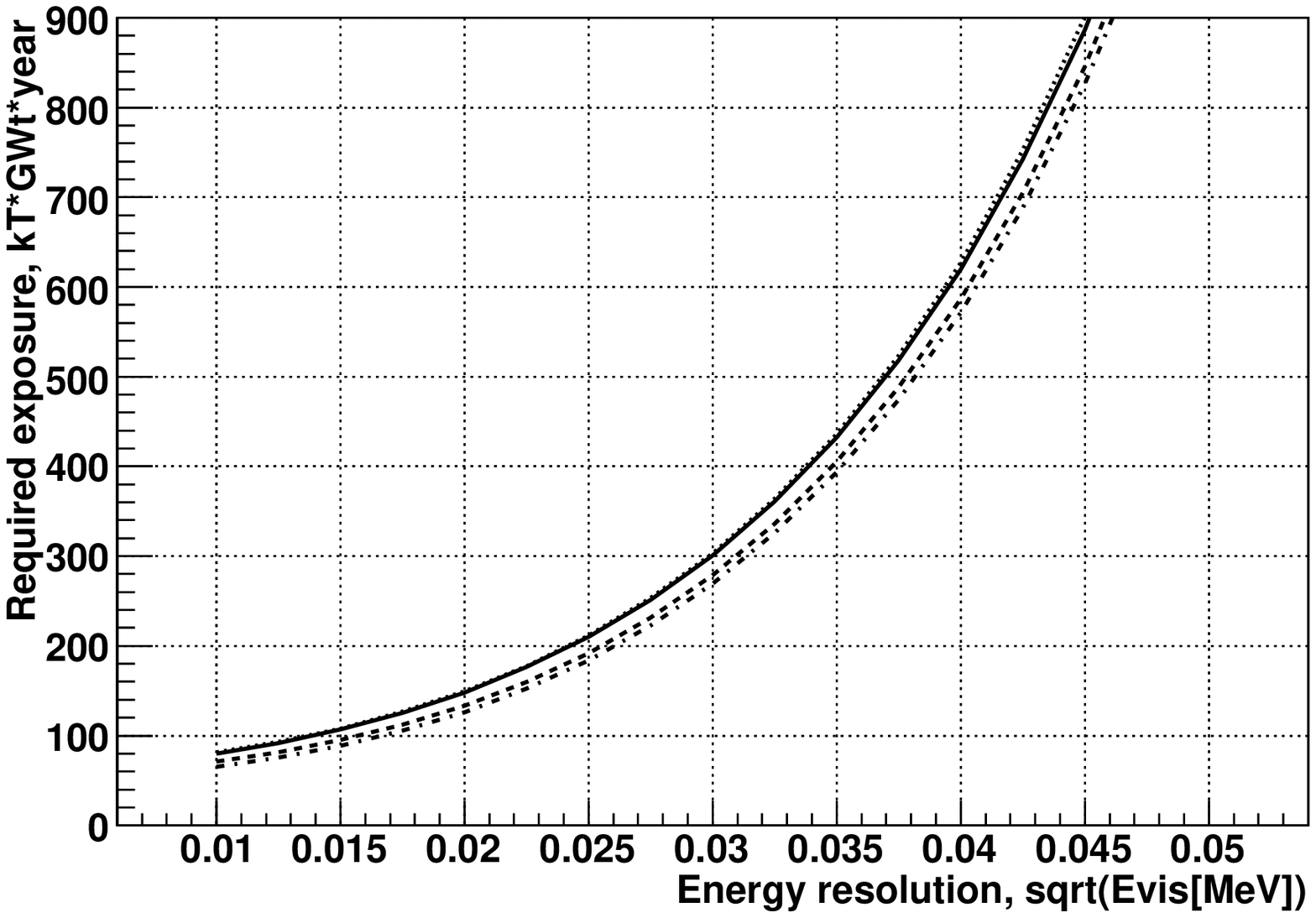}}
\caption{\textit {Assuming $\sin^22\theta_{13} = 0.05$:} 
exposure yielding the sensitivity of $0.025\times 10^{-3} eV^2$ in 
$\Delta m^2_{13}$ within a given mass hierarchy, 
as a function of detector energy resolution: with unconstrained 
detector systematics (dotted), with ``default'' detector systematics (solid), 
assuming no detector systematics (dashed), assuming no detector systematics 
and no geo-neutrinos (dot-dashed). 60 km baseline is assumed.}
\label{dms13R-50}
\end{figure}

\begin{figure}[p]
\centering
\scalebox{0.7}{\includegraphics{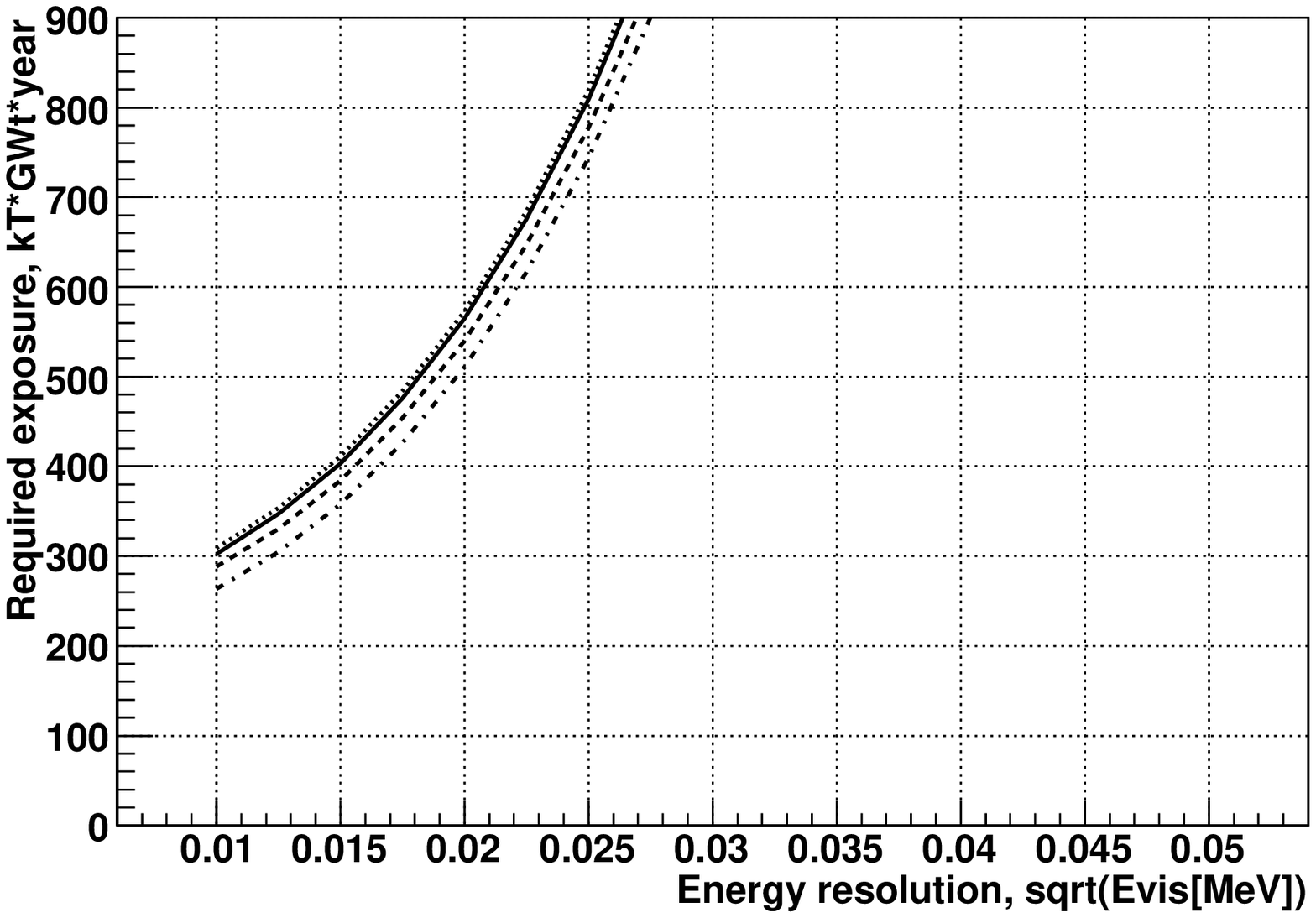}}
\caption{\textit {Assuming $\sin^22\theta_{13} = 0.025$:} 
exposure yielding the sensitivity of $0.025\times 10^{-3} eV^2$ in 
$\Delta m^2_{13}$ within a given mass hierarchy, 
as a function of detector energy resolution: with unconstrained 
detector systematics (dotted), with ``default'' detector systematics (solid), 
assuming no detector systematics (dashed), assuming no detector systematics 
and no geo-neutrinos (dot-dashed). 60 km baseline is assumed.}
\label{dms13R-25}
\end{figure}

Figure \ref{dms13-50} and \ref{dms13-25} show that shorter baselines are 
better for this study,
although this trend is not as pronounced as with $\theta_{13}$ measurement
and actually reverses below 25 km. It is clear that $\Delta m^2_{13}$ study
is not systematics-constrained, including the systematics
from geo-neutrinos. On the other hand, the dependence on energy resolution 
for this measurement
is even stronger than that for $\theta_{13}$ (Figure \ref{dms13R-50} and 
\ref{dms13R-25}).

\section{Neutrino mass hierarchy}

\begin{figure}[p]
\centering
\scalebox{0.7}{\includegraphics{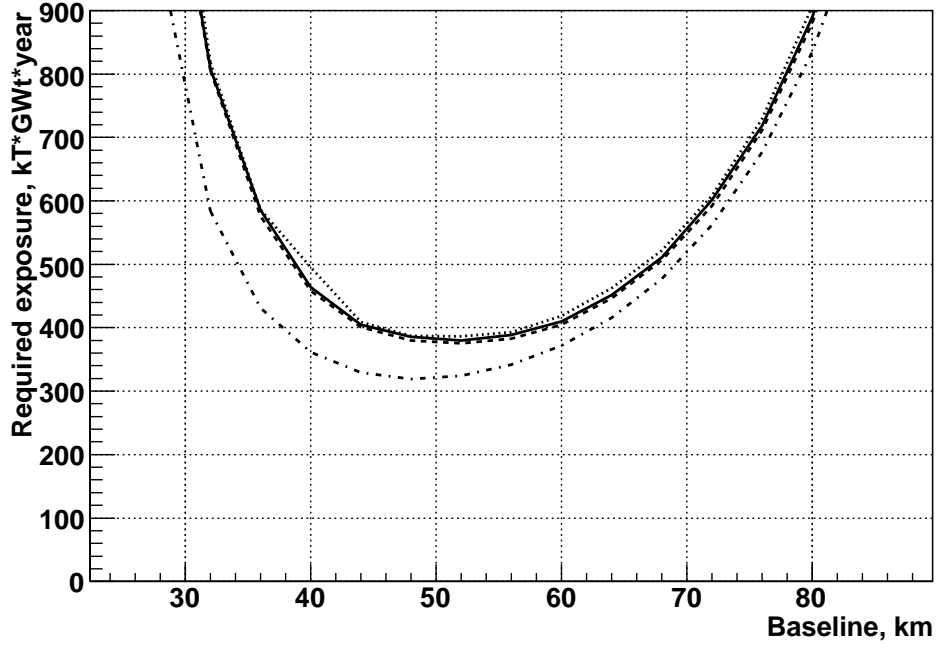}}
\caption{\textit {Assuming $\sin^22\theta_{13} = 0.05$:} 
exposure necessary to discriminate the neutrino mass hierarchies to 
66.8\% CL, as a function of baseline: with unconstrained 
detector systematics (dotted), with ``default'' detector systematics (solid), 
assuming no detector systematics (dashed), assuming no detector systematics 
and no geo-neutrinos (dot-dashed). Detector energy resolution equal to 
$0.025\times\sqrt{E_{vis}[MeV]}$ is assumed.}
\label{hier}
\end{figure}

Like the $\Delta m^2_{13}$/$\Delta m^2_{23}$ measurement, the hierarchy study 
depends on the actual value of $\theta_{13}$. Our calculations suggest that 
for any $\theta_{13}$ it takes more statistics to make a high confidence
level conclusion about the hierarchy than to measure $\Delta m^2_{13}$ and
$\Delta m^2_{23}$ to an accuracy of $0.025 \times 10^{-3}eV^2$. 
This makes
reliable hierarchy determination with a 10 kt detector feasible
only if the mixing angle turns out to be quite high ($\sin^22\theta_{12} \ge
0.05$). The sensitivity dependence on this value is approximately quadratic
in the exposure as well. The baseline profile
of the sensitivity to the hierarchy is shown in Figure \ref{hier}. After
taking into account the geo-neutrino background and the uncertainties the
optimum baseline remains in the same range as was found earlier with
less comprehensive models \cite{Hano1, Venice} --- 50 km or slightly more. 
Of the systematic errors, the most damaging is the geo-neutrino flux 
uncertainty, although its effect is not as decisive as for the
solar parameter studies. 

\begin{figure}[p]
\centering
\scalebox{0.7}{\includegraphics{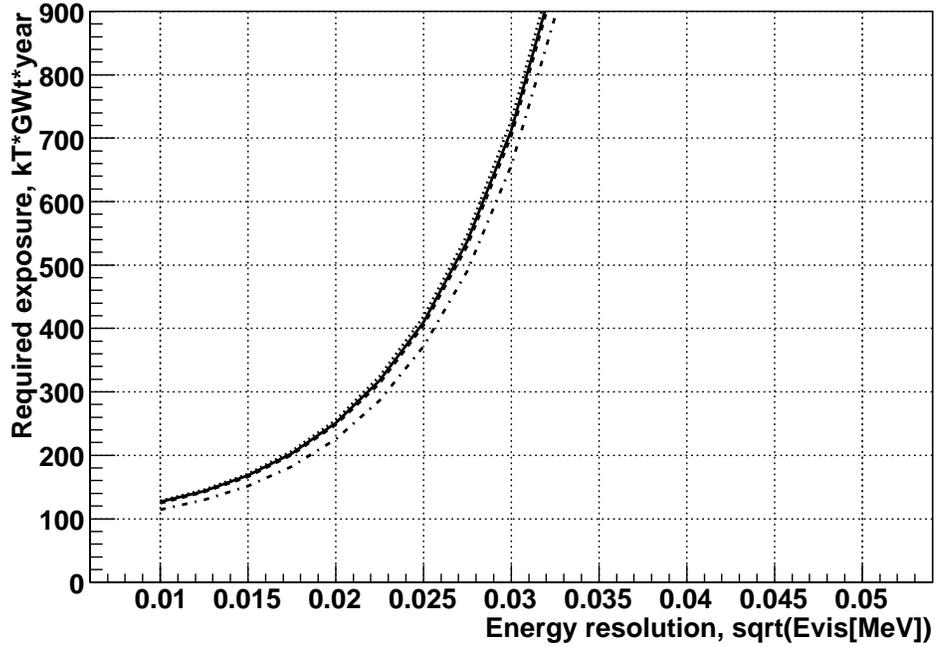}}
\caption{\textit {Assuming $\sin^22\theta_{13} = 0.05$:} 
exposure necessary to discriminate the neutrino mass hierarchies to 66.8\% CL,
as a function of detector energy resolution: with unconstrained 
detector systematics (dotted), with ``default'' detector systematics (solid), 
assuming no detector systematics (dashed), assuming no detector systematics 
and no geo-neutrinos (dot-dashed). 60 km baseline is assumed.}
\label{hier-R}
\end{figure}

The hierarchy study proves to be the most demanding of the detector energy
resolution (Figure \ref{hier-R}). Even within the best values for that 
parameter of the detector achievable today the sensitivity dependence on the 
energy resolution enters its asymptotic 4-th power curve. In particular, this
implies that between two detectors of the same photocathode area but 
different volumes, the bigger detector will offer \textit{inferior} 
sensitivity: all other parameters being equal, a smaller relative 
photocathode coverage will lead to lower resolution which will prevail 
over the higher reactor $\bar\nu_e$ statistics. 

Theoretically, the mass hierarchy study is secondary to
the measurement of $\Delta m^2_{13}$ and $\Delta m^2_{23}$ and is
determined immediately after those mass differences are found. The analysis
of the oscillated $\bar\nu_e$ energy spectrum yields all mass differences
(provided, $\theta_{13} \ne 0$ and $\theta_{12} \ne \pi/4$). In practice, 
however,the squared mass differences can 
be measured with limited accuracy only and at a limited CL. This may not
be sufficient to determine the hierarchy. Moreover, it has been found 
\footnotemark that for any combination of $\Delta m_{13}^2$ and 
$\Delta m_{23}^2$ there exists another one (denoted below as 
$\Delta' m_{13}^2$ and $\Delta' m_{23}^2$) that delivers a similar oscillation 
pattern despite comprising the oppositie mass hierarchy. The similarity is 
never perfect and, given enough statistics, it is always possible to 
distinguish between the two spectra but it may take
much more exposure to discriminate between those ``conjugate'' opposite 
hierarchy
solutions than to constrain the squared mass differences within one of
the solutions with a remarkable precision. 

\footnotetext {This problem had been pointed to in \cite {Petcov2} and 
\cite{SteveP}. Our sumulations confirm it, no matter whether Fourier 
transform is used in the data analysis or not.}

\begin{figure}[p]
\centering
\scalebox{0.44}{\includegraphics{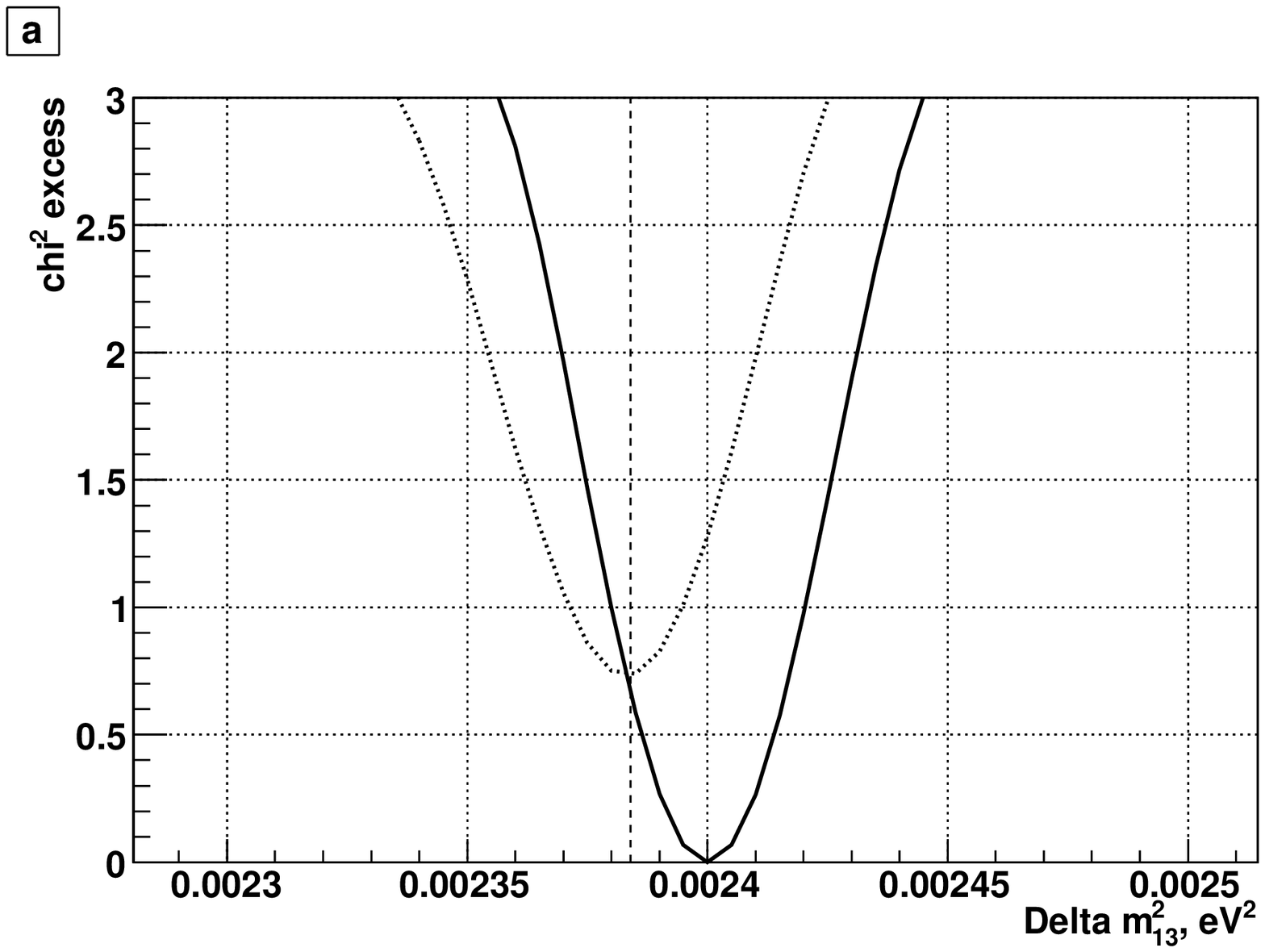}}
\scalebox{0.44}{\includegraphics{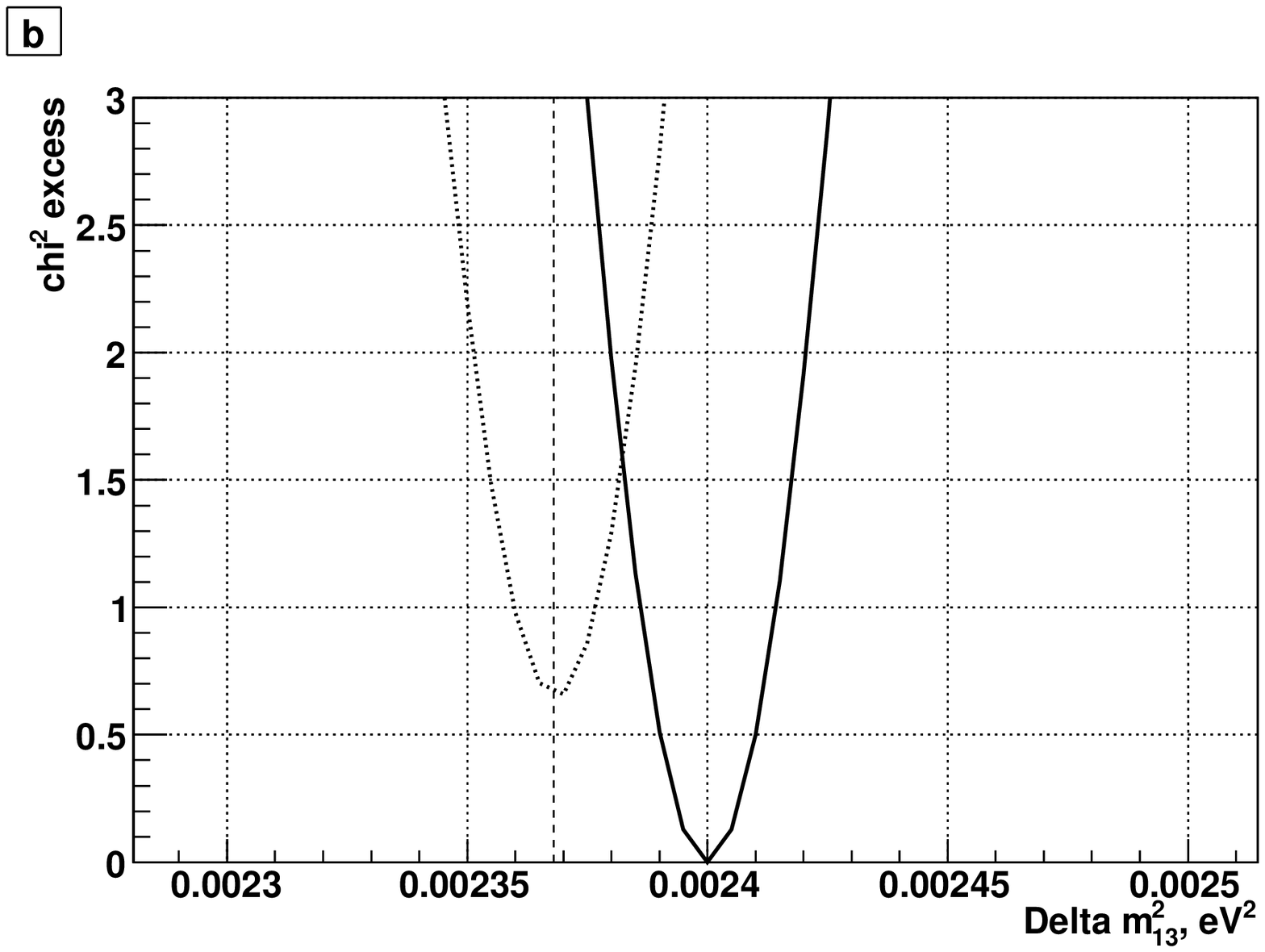}}
\scalebox{0.44}{\includegraphics{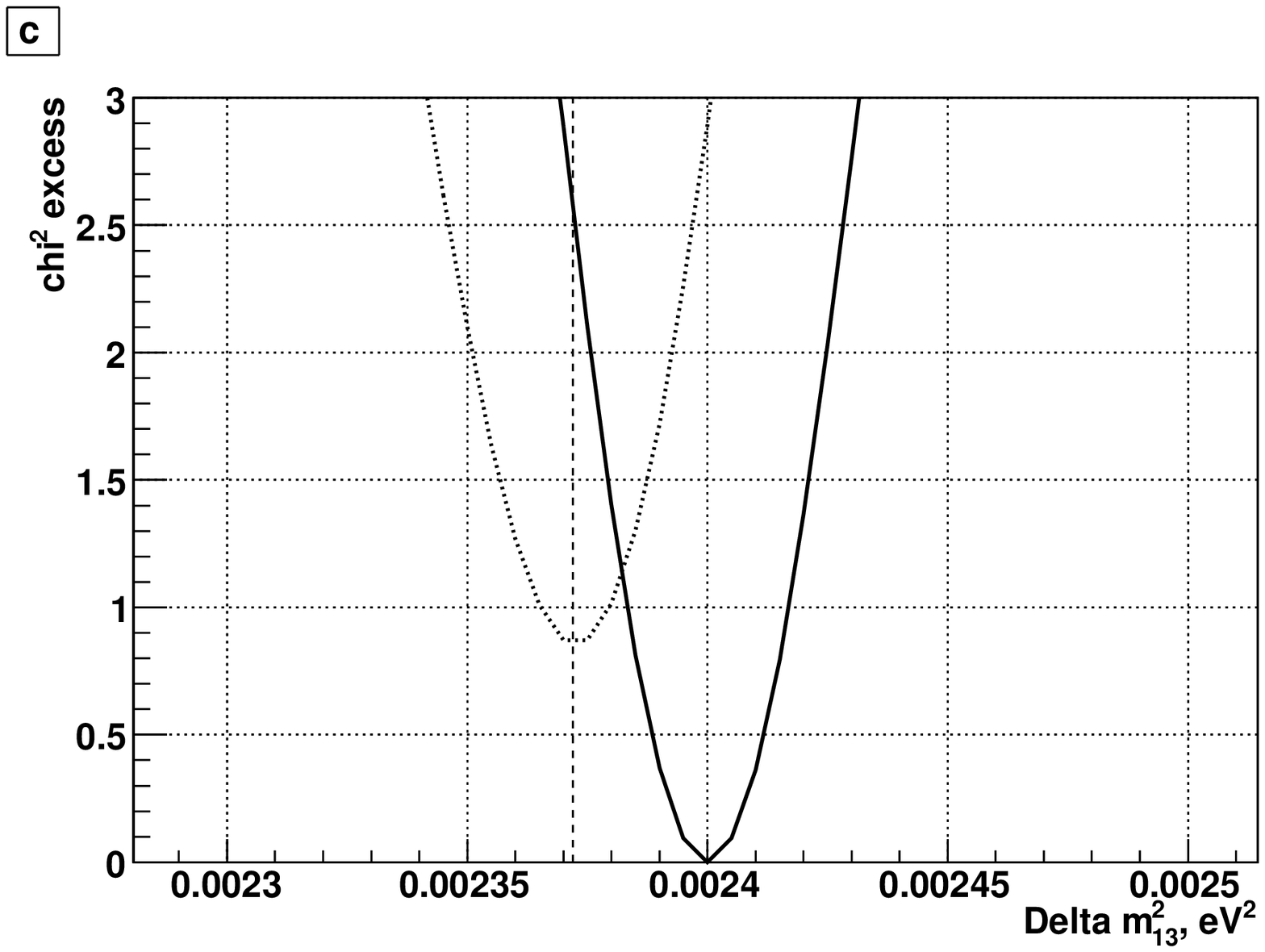}}
\scalebox{0.44}{\includegraphics{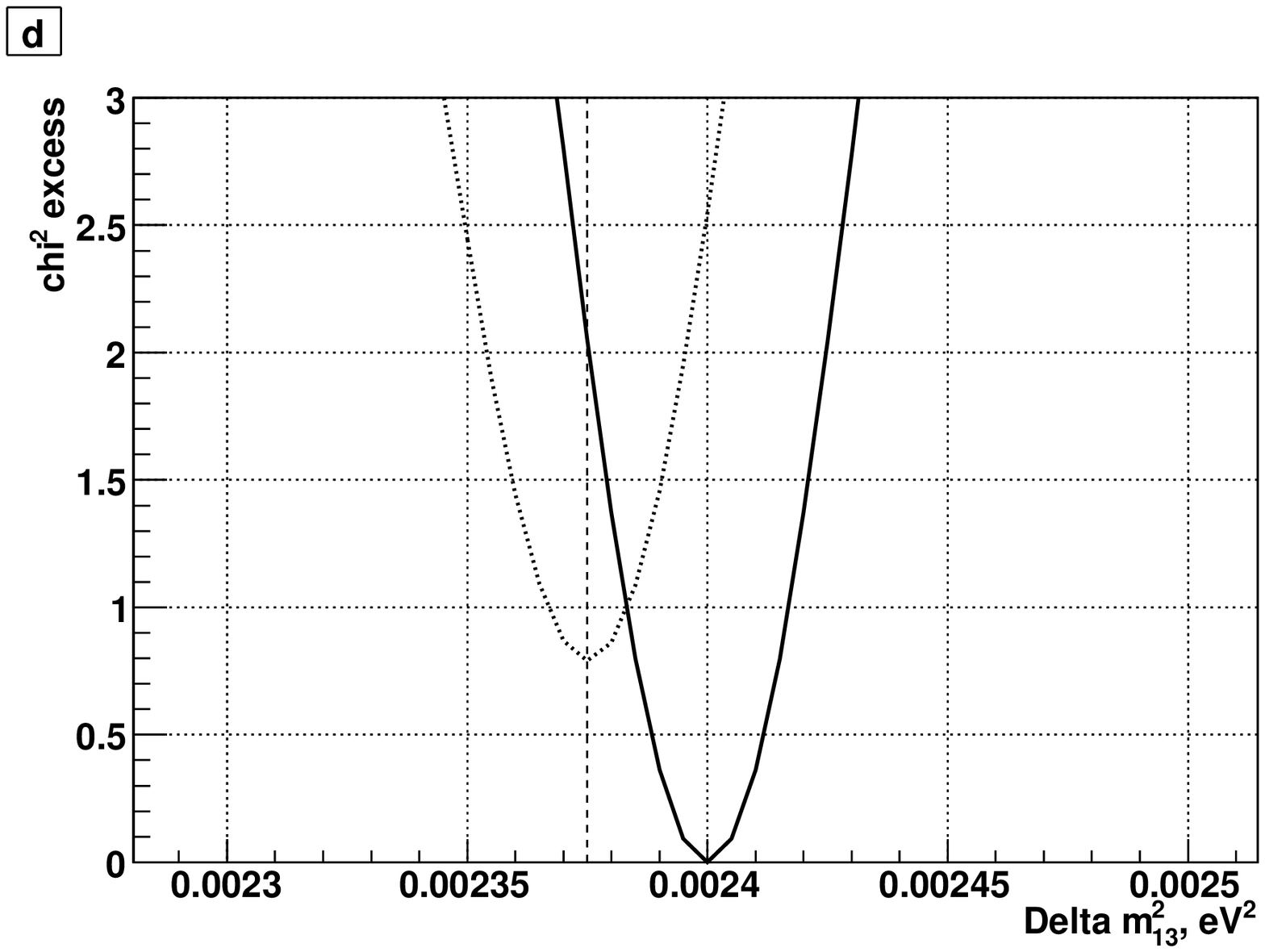}}
\caption{$\chi^2$ excess for an alternative hypothesis about the
$\Delta m^2_{13}$ and the mass hierarchy over the ``real'' one for
which the events were simulated ($\Delta m^2_{13} = 2.4\times10^{-3}eV^2$,
normal hierarchy). Solid line is for ``correct'' hierarchy, dotted -- for
the ``wrong'' one. The larger the $\chi^2$ excess, the easier the
alternative hypothesis to rule out. (a) single detector at 60 km;
(b) single detector at 40 km; (c) two half-sized detectors, one at
60 km and another at 40 km; (d) single detector at 50 km.
}
\label{likedef}
\end{figure}

In Figure \ref{likedef}, the curves provide the measure of relative
``unlikeliness'' of an alternative hypothesis, assuming 
normal hierarchy and $\Delta m^2_{13} = 2.4 \times 10^{-3}eV^2$,
for which the experiment was simulated. 
Zero $\chi^2$ means an indistinguishable hypothesis, the higher
its value, the less statistics is needed to discriminate the hypothesis. 
Solid lines
show the normal hierarchy and dotted lines show the inverted one. The dashed
vertical lines through the centers of the dotted curves point to the 
$\Delta' m^2_{13}$ which combined with the inverted hierarchy provide
the closest similarity to the simulated physical spectrum. 
Comparison of Figure \ref{likedef}(a) and Figure \ref{likedef}(b) explains
why the 60 km baseline offers better hierarchy discrimination than 40 km,
although the latter yields significantly better sensitivity to 
$\Delta m^2_{13}$/$\Delta m^2_{23}$ within each of the two hierarchies: the 
vertex of the quasi-parabolic dotted curve is located higher for 60 km. 

The ``conjugate'' $\Delta' m^2_{13}$ for a given real $\Delta' m^2_{13}$
is not the same for different baselines, which implies that two 
measurements at different baselines may offer improved efficiency. For 
instance, the allowable values of $\Delta' m^2_{13}$ to which the 60 km 
baselined measurement is least sensitive are much better excluded at 40 
km and vice versa. In case of a land-based detection, a multiple-detector
configuration can be considered. Hanohano, additionally, can use the 
advantage of its movability and make two consecutive exposures instead of one
twice as long. Indeed, as the comparison between Figure \ref{likedef}(c) 
and Figure \ref{likedef}(d) shows, the combination of 60 and 40 km
baselined observations should provide a better hierarchy resolution than 
one twice as long at the practically optimal 50 km. Although the
advantage is marginal, we considered only a two-baseline combination with
equal exposures. A more systematic optimization with different exposures
and possibly more than just two baselines should offer further gains.

\section{Conclusion}

%The new simulations confirm the point about the great potential of the 
%precision neutrino oscillation measurements with medium-baselined 
%reactor $\bar\nu_e$ experiments. Numerical estimations for the expected 
%accuracies are collected in Table \ref{table_conclusion}. One of the most
%compelling features of such an experiment is that the estimations for all
%the oscillation parameters can be performed in a single long-running 
%experiment at some fixed baseline. Although optimum baselines for
%different parameters are not the same, it turns out that the 55 km baseline
%is good enough for most or all of them (depending on the actual value
%of $\theta_{13}$) to dramatically improve on the accuracy compared to current
%best estimations. 

With a single detector and a 300 kiloton-gigawatt-year exposure and
a 5-6 GW thermal power $\bar\nu_e$ source, both $\Delta m^2_{12}$ and 
$\theta_{12}$ are expected to be measured with the accuracy of 1\%, a three
to five time improvement on the current best limits \cite{KamLAND3}. 
An experiment with a more powerful reactor 
$\bar\nu_e$ source but the same exposure due to a smaller detector size or
a shorter livetime will have a slight advantage because of the relatively 
smaller geo-neutrino effect.

Detector-associated systematic uncertainties do not appear to be a 
significant limiting factor in the resulting accuracy. These studies
are not particularly sensitive to the detector resolution and 
$5-6\%\times\sqrt{E_{vis}[MeV]}$ would be quite sufficient. However the 
expected sensitivity, especially that for $\Delta m^2_{12}$, will
be severely handicapped by the presence of geo-neutrinos and in particular by 
the lack of accurate estimation for the intensity of this background source, 
effectively turning its value into yet another systematic uncertainty 
--- the dominating one for this study. This is not a ``hard'' limitation since 
geo-neutrinos, having a different spectrum, can not mimic the 
reactor $\bar\nu_e$ oscillaton pattern, but rather an efficiency impairment. 
%Without this background and
%the associated systematic uncertainty the sensitivity would be several times
%times higher than we currently expect. 

The study of the $\theta_{13}$ mixing angle is different, in that the medium 
baselines that this study deals with are not optimal. On a per-event basis, the 
baseline dependence of sensitivity is rather flat, but short baselines gives
quadratically higher event rate and hence lower statistical error. On the other
hand, longer baselines offer more robustness with respect to the 
``efficiency'' systematic uncertainty. Given the careful design
of Hanohano or any similar (considering the size and the baseline) detector 
and its accurate calibration, it is still possible to 
reach an accuracy of better than 0.02 in $sin^22\theta_{13}$ evaluation, 
which is 
competitive with the dedicated experiments like Double Chooz and Daya Bay. 
The sensitivity to this mixing angle does not exhibit strong dependence on 
its own real value. In other words, setting the upper limit of 0.02 for
$\sin^22\theta_{13}$ if the angle happens to be zero takes about as much 
exposure as setting the range between 0.03 and 0.07 if the value is really 
0.05.

%A special short-baseline run to achieve a better accuracy for $\theta_{13}$
%is unlikely to be beneficial because there may not be a spot in
%the ocean close enough to a nuclear power plant and at the same time deep 
%enough to suppress the cosmic ray background which is neglected here. The 
%quantitative answer to this question deserves a dedicated study.

The measurement of $\Delta m^2_{13}$ and $\Delta m^2_{23}$ and the closely 
related question of neutrino mass hierarchy are common in their dependence 
on the actual value of $\theta_{13}$. If the angle turns out to be big enough 
(within currently allowed values), then some spectacular results are possible.
If it is zero or very small, nothing interesting can be measured 
with either Haonhano or any other similar experiment. The necessary exposure 
is approximately inversely quadratic to the value of $\sin^22\theta_{13}$
in both cases, although the hierarchy measurement generally requires much more
statistics. For a moderately optimistic
scenario in which $\theta_{13}=0.025$, Hanohano or a similar experiment can 
yield a very good estimation for the values of the squared mass differences 
but reliable mass hierarchy separation may call for prohibitively long 
exposures. 
Another feature that these measurements have in common is the requirement
for the excellent energy resolution. 

Generally, assuming either normal or inverted hierarchy, the problem of 
$\Delta m^2_{13}$ and $\Delta m^2_{23}$ becomes simpler, less demanding 
of the energy resolution and with higher chance of success
for unfavorably small values of $\theta_{13}$. Even
if the hierarchy question is not conclusively answered at that stage, 
the 
squared mass differences can still be measured, even though the remaining
hierarchy ambiguity will split the allowable solutions into two groups. In
such a case, the result of the $\Delta m^2_{13}$ and $\Delta m^2_{23}$
study will have the form:

\[
\Delta m^2_{13} = \Delta_{norm} \pm \Delta\delta_{norm}
\]
\[
\Delta m^2_{23} = \Delta_{norm}-\Delta m^2_{12} \pm \Delta\delta_{norm}
\]
for normal hierarchy, and
\[
\Delta m^2_{13} = \Delta_{inv} \pm \Delta\delta_{inv}
\]
\[
\Delta m^2_{23} = \Delta_{inv}+\Delta m^2_{12} \pm \Delta\delta_{inv}
\]
for inverted hierarchy, with one solution somewhat more favored over the 
other (i.e. should the hierarchy discrimination be achieved to marginal
confidence levels). Here $\Delta_{norm}$ and $\Delta_{inv}$ are best 
fit values
for the $\Delta m^2_{13}$ for the normal and inverted scenarios, respectively,
the $\Delta m^2_{12}$ is expected to be found from the same experiment with
superior accuracy and $\Delta\delta_{norm}$ and $\Delta\delta_{inv}$ are
the error bars for both the $\Delta m^2_{13}$ and $\Delta m^2_{23}$ in the 
normal and inverted hierarchy, respectively. 

The sensitivity properties for oscillation parameters in a single-baseline
experiment are summarized in Table \ref{SummaryTableDet} and 
\ref{SummaryTableNonDet}.

\begin{table}[h]
\caption[Summary Table]{Parameter sensitivity properties for
$\theta_{12}$, $\Delta m^2_{12}$ and $\theta_{13}$ with
Hanohano or similar detectors.}
\begin{center}
\begin{tabular}{|l||c|c|c|}
\hline
Parameter & $\theta_{12}$ & $\Delta m^2_{12}$ & $\theta_{13}$ \\
\hline
\hline
Detector systematics dependence & low & low & high \\
Geo $\bar\nu_e$ dependence & high & high & low \\
$\bar\nu_e$ energy resolution dependence & low & low & high \\
Optimal baseline for single detector, km & 60-70 & 70-80 & $<$20 \\
Expected sensitivity & 0.01 & $0.07\times 10^{-5}eV^2$ & 0.02  \\
\hline
\end{tabular}
\label{SummaryTableDet}
\end{center}
\end{table}

\begin{table}[h]
\caption[Summary Table]{Sensitivity properties for $\Delta m^2_{13}$
(or $\Delta m^2_{23}$) and neutrino mass hierarchy.}
\begin{center}
\begin{tabular}{|l||c|c|}
\hline
Parameter & $\Delta m^2_{13}$ & M. H. \\
\hline
\hline
Detector syststematics dependence & low & low \\
Geo $\bar\nu_e$ dependence & low & avg \\
$\bar\nu_e$ energy resolution dependence & v. high & extreme \\
Dependence on $\theta_{13}$ & yes & yes \\
Optimal baseline for single detector, km & $<$30 & 50 \\
\hline
\end{tabular}
\label{SummaryTableNonDet}
\end{center}
\end{table}

%\footnotetext[1]{ When dependence present, no robust sensitivity estimations
%can be made before $\theta_{13}$ is measured}
%\footnotetext[2]{ For a single detector at 60 km and
%assuming 300 gigawatt-kiloton-year exposure; for multiple detectors can
%be better}
%\footnotetext[3]{ in $sin^2(2\theta)$}

As was proposed in \cite{Grossman} and discussed in numerous later 
publications, neutrinos
may have non-standard interactions which could affect the flavor content at the
source and also the flavor content detected. For our case at hand, it means 
that the observed mixing angles $\theta_{12}$ and $\theta_{13}$ in general will
differ from the ``true'' mixing angles. For example, the measured 
$\theta_{13}$ can be larger than the true $\theta_{13}$ \cite{Ohlsson}. Since 
the survival probability (\ref{3flavor}) depends on the effective $\theta_{13}$,
the NSI have no adverse effect on the determination of $\Delta m^2_{13}$, 
$\Delta^2_{23}$ and the neutrino mass hierarchy. In fact an effective 
$\theta_{13}$ larger than the real one will be advantageous for these studies.

At the same time, the effective $\theta_{13}$ measured at different baselines
are going to be different, should these interactions take place. This 
way, medium baseline experiments targeting $\theta_{13}$ will become 
complementary to the short baseline ones in testing new physics.

The two most important qualitative conclusions from this study are the 
following:
\begin{itemize}
\item
  Medium-baselined $\bar\nu_e$ oscillation experiments are not
systematics-constrained. This follows from the shapes of oscillated
spectra. In particular, physically feasible systematic errors do
not tend to imitate the spectral distortions characteristic of the neutrino
oscillations.
\item
  There is no single baseline optimal for all oscillation studies. The
difference in sensitivity profiles is big enough to give an advantage
to multiple detector or/and movable detector configurations.
Even in individual studies where a pronounced baseline optimum exists, 
a multiple baseline configuration can outperform a single baseline 
configuration, as has been shown for the neutrino mass hierarchy 
discrimination case.
\end{itemize}

%After the ``fundamental physics'' phase of the experiment is over,
%the underwater detector can be moved to a more distant position for the
%dedicated, high precision study of terrestrial antineutrinos, although even
%with the reactor background considered here the terrestrial antineutrino flux
%will already have been evaluated to a much higher precision than possible 
%with current detectors.

\begin{acknowledgments}
This work was partially supported by the U.S. Department of Energy grant
DE-FG02-04ER41291 and the University of Hawaii.
\end{acknowledgments}


\begin{thebibliography}{99}


\bibitem{Pontecorvo}
  B. Pontecorvo, Sov. Phys. JETP \textbf{7}, 172 (1959).
\bibitem{MNS}
  Z. Maki, M. Nakagawa and S. Sakata, Prog. Theor. Phys. \textbf{28}, 870
(1962).

\bibitem{SNO3} 
  B. Aharmim \textit{et al.}, Phys. Rev. \textbf{C72} 055502 (2005).

\bibitem{KamLAND3} 
  S. Abe, \textit{et al.}, Phys. Rev. Lett. \textbf{100}, 221803 (2008).

\bibitem{ReinesCowan}
  F. Reines and C. L. Cowan, Jr., Phys. Rev. \textbf{92}, 830 (1953).
\bibitem{Davis}
  R. Davis, Jr., \textit{et al.}, Phys. Rev. Lett. \textbf{20}, 1205 (1968).

\bibitem{Hano0} S.T. Dye \textit{et al.},
%``Earth Radioactivity Measurements with a Deep Ocean Anti-neutrino
%Observatory,"
Earth, Moon, and Planets \textbf{99} (2006) 241-252.

\bibitem{Venice} J. G. Learned, S. T. Dye, S. Pakvasa in 
\textit{Twelfth International Workshop on Neutrino Telescopes}, 
ed. Baldo Ceolin (Instituto Veneto di Scienze, Padova), p 235 (2007).

\bibitem{Wolfenstein}
  L. Wolfenstein, Phys. Rev. D \textbf{17}, 2369 (1978).
\bibitem{MikheevSmirnov}
  S. P. Mikheev and A. Yu. Smirnov, Sov. J. Nucl. Phys. \textbf{42}, 913
(1985).

\bibitem{Schlattl}
  H. Schlattl, 
%``Can three-flavor oscillations solve the solar neutrino problem?'' 
Phys. Rev. \textbf{D64}, 013009 (2001).

\bibitem{Bilenky} S. M.  Bilenky ,  D. Nicolo and  S. T. Petcov,
%``Constraints on $|U_{e3}|^2$ from a Three-Neutrino Oscillation Analysis 
%of the CHOOZ Data'', 
Phys. Lett. \textbf{B538} (2002) 77-86.

\bibitem{SNO}
  Q. R. Ahmad, \textit{et al.}, Phys. Rev. Lett. \textbf{89} 011301 (2002).

\bibitem{SK-solar} S. Fukuda \textit{et al.}, Phys. Lett. \textbf{B539}
  (2002) 179-187.

\bibitem{SuperK2004}
  M. Ishitsuka, for the Super-Kamiokande Collaboration, hep-ex/0406076 (2004).
\bibitem{K2K}
  M. H. Ahn, \textit{et. al.}, Phys. Rev. \textbf{D74}, 072003 (2006).
\bibitem{MINOS}
  P. Adamson \textit{et al. (MINOS)}, Phys. Rev. \textbf{D73} (2006) 072002.

\bibitem{ItalianSpanish}
  G. L. Fogli, \textit{et. al.}, Phys. Rev. Lett. \textbf{101} 141801 (2008).


\bibitem{Schreckenbach}
  K. Shreckenbach \textit{et al.}, Phys. Lett. \textbf{B160} 325 (1985).
\bibitem{VogelReactor}
  P. Vogel \textit{et al.}, Phys. Rev. \textbf{C24} 1543 (1981).
\bibitem{Vogel}
P. Vogel, J. F. Beacom, Phys. Rev. D \textbf{60}, 053003 (1999).

\bibitem{Petcov1}  S. T. Petcov and M. Piai, Phys. Lett B. \textbf{533},
94 (2002) arXiv:hep-ph/0112074.
\bibitem{Petcov2} S. Choubey, S. T. Petcov, M. Piai, Phys. Rev. D \textbf{68}, 
113006 (2003) arXiv:hep-ph/0306017v1.
\bibitem{Petcov3} A. Bandyopadhyay, S. Choubey, S. Goswami, 
S.T. Petcov, D.P. Roy,
% ``Neutrino Oscillation Parameters After High Statistics KamLAND Results'',
(2008) arXiv:0804.4857v1 [hep-ph]

\bibitem{Borexino} M. Balata, \textit{et al.}, Eur.Phys.J. \textbf{C47} 
  (2006) 21-30.

\bibitem{KLGeo}
  T. Araki \textit{et al.}, 
% ``Experimental investigation of geologically produced antineutrinos with 
% KamLAND''
, Nature \textbf{436}, 499 (2005).


\bibitem{Hano1} J. G. Learned, S. T. Dye, S. Pakvasa, R. C. Svoboda, 
%``Determination of Neutrino Mass Hierarchy and $\theta_{13}$ with a 
%Remote Detector of Reactor Antineutrinos'', 
hep-ex/0612022, Phys. Rev. D (in press).



\bibitem{KamLAND1}
  K. Eguchi \textit{et al.}, Phys. Rev. Lett. \textbf{90}, 021802 (2003).
\bibitem{KamLAND2}
  T. Araki \textit{et al.}, Phys. Rev. Lett. \textbf{94}, 081801 (2005).

%\bibitem{SNO2}
%  S. N. Ahmed \textit{et al.}, Phys. Rev. Lett. \textbf{92}, 181301 (2004).
\bibitem{CHOOZ}
  M. Apollonio \textit{et al.}, Eur. Phys. J. C \textbf{27}, 331 (2003).


\bibitem{GlenCowan} Glen Cowan, ``Statistical Data Analysis'', Oxford Science
Publications, 1998.

\bibitem{SteveP} Stephen Parke, private communication.

\bibitem{Grossman} Y. Grossman, Phys Lett \textbf{B359} (1995) 141.

\bibitem{Ohlsson} T. Ohlsson and H. Zhang, arXiv:0809.4835 (2008) [hep-ph].

\end{thebibliography}
\end{document}